\documentclass[12pt]{article}
\pdfoutput=1
\usepackage{mathrsfs}
\usepackage{graphicx}
\usepackage{xr-hyper} 
\usepackage{xcolor} 
\usepackage{geometry} 
\usepackage{amsmath} 
\usepackage{amssymb} 
\usepackage[hyphens]{url}
\usepackage[hypertexnames=false]{hyperref}
\usepackage[title,toc]{appendix}
\newcommand{\georddtitle}{
    A Bayesian Nonparametric Approach to Geographic Regression Discontinuity Designs:
    Do School Districts Affect NYC House Prices?
}
\hypersetup{
    colorlinks=true,
    pdfborder=0 0 1,
    pdftitle=\georddtitle{},
    allcolors=[rgb]{0.00,0.38,0.59},
    allbordercolors=[rgb]{0.60,0.76,0.86}
}
\usepackage{natbib}
\usepackage{bm}
\usepackage{mathtools}
\usepackage{pdflscape}
\usepackage[nodisplayskipstretch]{setspace}
\renewcommand*{\arraystretch}{0.8} 
\def\SmallColSep{\setlength{\arraycolsep}{3pt}}

\usepackage{enumitem}
\newlist{flatlist}{enumerate*}{1}
\setlist[flatlist]{label=(\arabic*)}

\newcommand{\mathbold}[1]{\bm{#1}} 

\DeclarePairedDelimiter{\parenthesis}{\lparen}{\rparen}
\DeclarePairedDelimiter{\squarebracket}{\lbrack}{\rbrack}
\DeclarePairedDelimiter{\curlybracket}{\lbrace}{\rbrace}
\DeclarePairedDelimiter{\absolutevalue}{\lvert}{\rvert}
\newcommand{\del}[1]{\parenthesis*{#1}}
\newcommand{\sbr}[1]{\squarebracket*{#1}}
\newcommand{\cbr}[1]{\curlybracket*{#1}}
\newcommand{\abs}[1]{\absolutevalue*{#1}}
\newcommand*{\diffdchar}{d}
\newcommand*{\dif}[1]{\mathop{\diffdchar #1}}

\let\Pr\relax
\DeclareMathOperator{\Pr}{\mathbb{P}}
\DeclareMathOperator{\E}{\mathbb{E}}

\DeclareMathOperator{\cov}{{Cov}}
\DeclareMathOperator{\var}{{var}}
\DeclareMathOperator{\Ind}{\mathbb{I}}

\DeclareMathOperator{\normal}{\mathcal{N}}

\DeclareMathOperator{\ones}{\mathbf{1}}
\DeclareMathOperator{\GP}{\mathcal{GP}}
\newcommand{\building}{\mathtt{BuildClass}}
\newcommand{\district}{\mathtt{Distr}}

\newcommand*{\trans}{^{\intercal}}

\newcommand*{\area}{\mathcal{A}}
\newcommand*{\treat}{T}
\newcommand*{\ctrol}{C}
\newcommand*{\treatind}{Z}
\newcommand*{\treatarea}{\area{}_{\treat}}
\newcommand*{\ctrolarea}{\area{}_{\ctrol}}

\newcommand*{\sigmaf}{\sigma_{\mathrm{GP}}}
\newcommand*{\sigman}{\sigma_{\epsilon}}
\newcommand*{\sigmabeta}{\sigma_{\beta}}
\newcommand*{\sigmamu}{\sigma_{m}}
\newcommand*{\svec}{\mathbold{s}}
\newcommand*{\dvec}{\mathbold{d}}
\newcommand*{\wvec}{\mathbold{w}}
\newcommand*{\yvec}{\mathbold{y}}
\newcommand*{\Yvec}{\mathbold{Y}}
\newcommand*{\yt}{\Yvec_{\treat}}
\newcommand*{\yc}{\Yvec_{\ctrol}}

\newcommand*{\muvec}{\mathbold{\mu}}
\newcommand*{\betavec}{\mathbold{\beta}}
\newcommand*{\residvec}{\mathbold{R}}
\newcommand*{\indep}{\protect\mathpalette{\protect\independenT}{\perp}}
\def\independenT#1#2{\mathrel{\rlap{$#1#2$}\mkern2mu{#1#2}}}

\newcommand*{\border}{\mathcal{B}}
\newcommand*{\sentinel}{\mathbold{b}}
\newcommand*{\numsent}{R}
\newcommand*{\sentinels}{\sentinel_{1:\numsent}}
\newcommand*{\isent}{r}
\newcommand*{\sentinelset}{\cbr{\sentinel_1,\ldots,\sentinel_\numsent}}

\newcommand*{\eye}{\mathbf{I}}

\newcommand*{\tauw}{\tau^{w}}
\newcommand*{\unifavg}{\tau^{\mathrm{UNIF}}}
\newcommand*{\invvar}{\tau^{\mathrm{INV}}}
\newcommand*{\taurho}{\tau^{\rho}}
\newcommand*{\tauproj}{\tau^{\mathrm{PROJ}}}
\newcommand*{\taugeo}{\tau^{\mathrm{GEO}}}
\newcommand*{\taupop}{\tau^{\mathrm{POP}}}

\newcommand*{\modnull}{\mathscr{M}_0}
\newcommand*{\modalt}{\mathscr{M}_1}
\newcommand*{\degree}{{\,^\circ}}

\DeclareMathOperator{\proj}{proj}
\DeclareMathOperator{\dist}{dist}
\newcommand*{\buffer}{\Delta}
\newcommand*{\vicinity}[1]{\Ind^\buffer\del{#1}}
\newcommand*{\hyperparam}{\bm{\theta}}

\newcommand*{\taubold}{\bm{\tau}}
\newcommand*{\weightb}{w_{\border}}
\newcommand*{\wt}{\wvec_{\treat}}   
\newcommand*{\wc}{\wvec_{\ctrol}}
\newcommand*{\gridres}{\nu}
\newcommand*{\grid}{G^\gridres}
\newcommand*{\Dmat}{\mathbold{D}}
\newcommand*{\Kmat}{\mathbold{K}}
\newcommand*{\Xmat}{\mathbold{X}}
\newcommand*{\Wmat}{\mathbold{W}}
\newcommand*{\SigmaMat}{\mathbold{\Sigma}}
\newcommand*{\KBB}{\Kmat_{\border \border}}
\newcommand*{\KBT}{\Kmat_{\border \treat}}
\newcommand*{\KBC}{\Kmat_{\border \ctrol}}
\newcommand*{\STT}{\SigmaMat_{\treat \treat}}
\newcommand*{\SCC}{\SigmaMat_{\ctrol \ctrol}}
\newcommand*{\KTT}{\Kmat_{\treat \treat}}
\newcommand*{\KCC}{\Kmat_{\ctrol \ctrol}}
\newcommand*{\KTC}{\Kmat_{\treat \ctrol}}
\newcommand*{\WT}{\Wmat_{\treat}}
\newcommand*{\WC}{\Wmat_{\ctrol}}

\def\equationautorefname~#1\null{(#1)\null}

\newcommand{\georddkeywords}{Gaussian processes; kriging; bayesian testing; causal inference; regression discontinuity; treatment effect; housing market}
\hypersetup{pdfkeywords=\georddkeywords{}}
\hypersetup{pdfauthor=Maxime Rischard}

\usepackage{authblk}
\newcommand{\autorefexternal}[1]{\autoref{#1}}

\begin{document}
\singlespacing

\title{
    \georddtitle
}
\author[a]{Maxime Rischard
	\thanks{
		This research was supported by the National Science Foundation Graduate Research Fellowship Program under Grant No. 1144152, by the National Science Foundation under Grant No. 1461435, by DARPA under Grant No. FA8750-14-2-0117, by ARO under Grant No. W911NF- 15-1-0172, and by NSERC. Any opinions, findings, and conclusions or recommendations expressed in this material are those of the authors and do not necessarily reflect the views of the National Science Foundation, DARPA, ARO, or NSERC.
	}
}
\author[a]{Zach Branson}
\author[b]{Luke Miratrix}
\author[c]{Luke Bornn}
\affil[a]{Department of Statistics, Harvard University}
\affil[b]{Graduate School of Education, Harvard University}
\affil[c]{Simon Fraser University}
\maketitle

\begin{abstract}
    Most research on regression discontinuity designs (RDDs) has focused on univariate cases, where only those units with a ``forcing'' variable on one side of a threshold value receive a treatment.
    Geographical regression discontinuity designs (GeoRDDs) extend the RDD to multivariate settings with spatial forcing variables.
    We propose a framework for analysing GeoRDDs, which we implement using Gaussian process regression. 
    This yields a Bayesian posterior distribution of the treatment effect at every point along the border.
    We address nuances of having a functional estimand defined on a border with potentially intricate topology, particularly when defining and estimating causal estimands of the local average treatment effect (LATE).
    The Bayesian estimate of the LATE can also be used as a test statistic
    in a hypothesis test with good frequentist properties, 
    which we validate using simulations and placebo tests.
    We demonstrate our methodology with a dataset of property sales in New York City,
    to assess whether there is a discontinuity in housing prices at the border between two school district.
    We find a statistically significant difference in price across the border between the districts with \(p\)=0.002, and estimate a 20\%  higher price on average for a house on the more desirable side.
\end{abstract}

\noindent%
\vfill
\newpage

\section{Introduction}\label{introduction}

Regression discontinuity designs (RDDs) are natural experiments characterized by the treatment assignment being fully determined by some covariates, which are termed ``forcing'' variables.
A typical RDD scenario arises when a treatment is given to all units with a forcing variable that falls below (or above) an arbitrary threshold value, and is withheld from units on the other side of the threshold.
If, as is often the case, the forcing variable is also predictive of the outcome of interest, then treatment assignment and outcomes are confounded, but by focusing on units near the threshold, a causal treatment effect can nonetheless be estimated.
The theory and methods for RDDs date from the 1960s, starting with \cite{thistlethwaite1960regression}.
\cite{cook2008waiting} trace the history of how interest in RDDs subsequently waned, until the late 1990s when the design saw renewed attention, theoretical progress, and applications in the social sciences.
More recently, beginning with \cite{papay2011extending}, methods have been developed to analyse RDDs with multiple forcing variables.
\cite{imbens2011regression} extend the local linear regression methods \citep[see][]{imbensrdd} that are popular for analysing RDDs with a single forcing variable (1D RDDs) to settings with multiple forcing variables.

Geographical regression discontinuity designs (GeoRDDs) are such RDDs where the forcing variables are spatial,
meaning that units within a certain region are assigned to treatment, while units in a neighboring region are assigned to control.
For example, in \cite{macdonald2015effect}, a private police force patrols a neighborhood, but stays out of surrounding areas, and a causal effect on crime rates is sought.
In \cite{chen2013evidence}, a policy applies south of the Huai River in China but not in the north, and pollution levels and life expectancies are measured to infer environmental and health impacts of the policy.
In our application, we seek to estimate the effect of school districts on house prices in New York City.

Practitioners often wish to use the well-established methods and software developed for 1D~RDDs with their spatial data.
It is therefore tempting to reduce a GeoRDD problem to a 1D~RDD by using the signed distance from the boundary (positive for treatment and negative for control) as the forcing variable, a method that we refer to as ``projected 1D~RDD,'' and which is used by both examples cited above.
However, this method can fail to capture the spatial variation in the outcomes, resulting in spatial confoundedness of the estimator.
We demonstrate the resulting bias in a simple example in \autorefexternal{sec:confounding} of the Supplementary Materials.
See also Section~4.2 of \cite{keele_titiunik_2015} for a discussion of this issue.

A more principled treatment of the GeoRDD is offered by \cite{keele_titiunik_2015}, who build theoretical foundations for the analysis of GeoRDDs.
They extend the identification assumptions that were formalized by \cite{hahn2001identification} for 1D~RDDs, and \cite{imbens2011regression} for multivariate RDDs, to GeoRDDs.
The main requirement for identification of the treatment effect is continuity of the conditional regression functions near the border.
Notably, this is violated if units can sort around the border, crossing the border to seek or avoid the treatment, a particular concern in GeoRDDs.
\cite{keele_titiunik_2015} discuss further pitfalls of GeoRDDs, such as the issue of compound treatments---when a geographical border determines the assignment of the treatment of interest, but also of other differences.

Several methods for estimating the treatment effect in GeoRDDs have been proposed.
\cite{keele_titiunik_2015} and \cite{keeleoverview} estimate the treatment effect using a modification of the projected 1D~RDD method by applying it locally around each point along the border, thus alleviating the problem of spatial confounding.
This is slightly different from the method of \cite{imbens2011regression} developed for multivariate regression discontinuities that are not necessarily geographical: they use a multivariate local linear regression to estimate the treatment effect at each boundary point, thus avoiding the need to specify a distance metric in the space of covariates.
\cite{keele2015enhancing} propose an alternative approach: they deploy the matching methods of \cite{zubizarreta2012using} to match units on opposite sides of the border that are near each other geographically and in other covariates, and then analyze the matched outcomes as if they came from a randomized experiment.
Their method requires the analyst to choose a ``buffer'' distance from the border, so that within this distance spatial variation can plausibly be assumed to be negligible, resulting in estimates that are robust to spatial confounding.

\label{sec:framework}

In this paper, we propose a framework for analysing GeoRDDs that is a spatial analogue of 1D~RDD methods.
Broadly, 1D~RDD methodologies are composed of three steps:
\begin{flatlist}
    \item
        fit a smooth \emph{function} to the outcomes against the forcing variable on each side of the threshold,
    \item
        extrapolate the functions to the \emph{threshold point}, and
    \item
        take the difference between the two extrapolations to estimate the treatment effect at the threshold point.
\end{flatlist}
Reusing the same methodological skeleton and applying it to geographical RDDs, our framework proceeds analogously:
\begin{flatlist}
    \item
        fit a smooth \emph{surface} to the outcomes against the geographical covariates in each region,
    \item
        extrapolate the surfaces to the \emph{border curve}, and
    \item
        take the \emph{pointwise} difference between the two extrapolations to estimate the treatment effect along the border.
\end{flatlist}

\cite{Branson:2017qy} proposed a Gaussian process regression (GPR) methodology that exhibits promising coverage and MSE properties compared to local linear regression for 1D~RDDs.
We believe this approach to be particularly suitable to GeoRDDs, as GPR is a well-established tool in spatial statistics (where it is known as kriging) for fitting smoothly varying spatial processes.
See \cite{banerjee2014hierarchical} for a textbook introduction to kriging for spatial data, and \cite{rasmussen2006gaussian} for a machine learning perspective.

In \autoref{sec:geordd_model}, we use GPR to estimate the treatment effect along the border by extending the model of \cite{Branson:2017qy} to geographical settings.
A peculiarity of GeoRDDs is that the estimand is a function defined everywhere along the border, which is a one-dimensional manifold embedded in two-dimensional space.
Furthermore, geographical borders, whether they be political or natural, are rarely simple straight lines.
The topology of borders complicates the definition and interpretation of estimands for the local average treatment effect (LATE), which we address in \autoref{sec:ate}.
We obtain Bayesian estimators for multiple possible LATE estimands and discuss their properties.
In \autoref{sec:hypothesis_testing} we turn to hypothesis testing, and propose a method to test against the null hypothesis of no treatment effect along the border.

In \autoref{sec:NYC_example}, we apply our methodology to a publicly available dataset of property sales in NYC to determine whether school districts affect property prices.
Focusing on a single border between two school districs, we estimate the treatment effect everywhere along the border, obtain estimates of the LATE, and perform and validate a hypothesis test.
We find a statistically significant difference in price across the border with a \(p\)-value of 0.002, and estimate that the same house located near the border will on average fetch an almost 20\% higher price in district 27 than in district 19.
However, this effect can not be attributed solely to the reputation of the school district, as this border also separates the boroughs of Brooklyn and Queens, thus confounding the causal effect of the districts.

\section{GeoRDD Modeling with Gaussian processes}
\label{sec:geordd_model}

We largely adopt the setup and notation for GeoRDDs laid out in \cite{keele_titiunik_2015}.
The outcomes \(Y_i\) of \(n\) units with spatial coordinates \(\svec_i\) are observed within an area \(\area\) of 2-dimensional coordinate space.
The units are separated into \(n_\treat\) treatment units in area \(\treatarea \subset \area\)
and \(n_\ctrol\) units in the control area \(\ctrolarea\).
The defining characteristic of GeoRDDs is that the two areas are adjacent but non-overlapping, intersecting only at the border \(\border\) between them.
Throughout this paper, points on the border are denoted by \(\sentinel\).
Under the potential outcomes framework for causal inference, each unit \(i\) has potential outcomes \(Y_{i\treat}\) and \(Y_{i\ctrol}\) under treatment and control respectively.
Let \(\treatind_i\) denote the treatment indicator, which is equal to one if unit \(i\) is in the treatment area, and zero if it is in the control area.
Unlike traditional randomized experiments, treatment assignment is a deterministic function of a unit's geographical coordinates \(\svec_i\): \(\treatind_i = \Ind\cbr{\svec_i \in \treatarea}\).
The observed outcome for unit \(i\) is \(Y_i = \treatind_i Y_{i\treat} + (1 - \treatind_i) Y_{i\ctrol}\).
We denote the vector of observed outcomes of the treatment units and control units respectively by \(\yt\) and \(\yc\), and \(\Yvec\) the vector formed by concatening \(\yt\) and \(\yc\).

For 1D RDDs, because the treatment and control regions do not overlap, the treatment effect is typically only inferred at the threshold \(X=b\).
As was already recognized by \cite{thistlethwaite1960regression}, this choice requires the least extrapolation of the fitted regression functions, which makes the estimated treatment more credible.
The estimand at the threshold can be obtained as the difference of the two limits of the expectation of the conditional regression functions
\begin{equation}
    \tau = \E\sbr{Y_{i\treat} \mid X_i=b} - \E\sbr{Y_{i\ctrol} \mid X_i=b} = \lim_{x \downarrow b} \E\sbr{Y \mid X=x} - \lim_{x \uparrow b} \E\sbr{Y \mid X=x}\,,
    \label{eq:rdd_univ_estimand}
\end{equation}
where the second equality requires the assumption that the conditional regression functions \(\E\sbr{Y_{i\treat} \mid X_i=x}\) and \(\E\sbr{Y_{i\ctrol} \mid X_i=x}\) are continuous in \(x\) (see Assumption 2.1 in \cite{imbensrdd} and the discussion that follows).
Analogously, we focus on the treatment effect at the border \(\border\) between the treatment and control regions:
\begin{equation}
\tau\colon \border \rightarrow \mathbb{R}
\quad\text{defined as}\quad
\tau\del{\sentinel} = \E\sbr{Y_{i\treat} - Y_{i\ctrol} \mid \svec_i = \sentinel}
\,.
\label{eq:functional_estimand}
\end{equation}
This is the functional estimand defined in \cite{imbens2011regression} and \cite{keele_titiunik_2015}.
For any \(\sentinel \in \border\), \(\tau(\sentinel)\) can be obtained as the difference of the two limits of the expected outcomes, approaching \(\sentinel\) from the treatment or the control side of the border, given the assumption that the conditional regression functions \(\E\sbr{Y_{i\treat} \mid \svec_i=\svec}\) and \(\E\sbr{Y_{i\ctrol} \mid \svec_i=\svec}\) are continuous in \(\svec\) within \(\area\).
This result is formalized under Assumption 2.2.2 by \cite{imbens2011regression} and Assumption 1 in \cite{keele_titiunik_2015}.

For computational reasons, we often represent the border as a set \(\sentinels=\sentinelset\), \(\sentinel_\isent \in \border\) of \(\numsent\) ``sentinel points'' along the border.
We denote by \(\tau(\sentinels)\) the \(\numsent\)-vector with \(r^\mathrm{th}\) entry \(\tau(\sentinel_r)\) of the treatment effect evaluated at \(\sentinel_r\).

\subsection{Model Specification}
\label{sec:twogp}
Our GeoRDD framework allows any method to be used to fit the outcomes on either side of the border.
In this paper we use Gaussian process regression (GPR) for this purpose.
GPR, known as kriging in the spatial statistics literature, is a Bayesian non-parametric method for fitting smooth functions. 
Recently, \cite{Branson:2017qy} showed GPR to be a promising approach for the analysis 1D RDDs.
Further inspired by the popularity of GPR in spatial statistics, we extend the model of \cite{Branson:2017qy} to geographical RDDs.

On each side of the border, we model the observed outcomes \(Y_i\) at location \(\svec_i\) as the sum of an intercept \(m\), a spatial Gaussian process \(f(\svec)\), and iid normal noise \(\epsilon\).
The Gaussian process has zero mean, and its covariance function is a modeling choice.
There is a rich literature of possible covariance functions, known as ``kernels'' in machine learning; see
\cite{banerjee2014hierarchical} and \cite{rasmussen2006gaussian} for examples.
In this paper we use the squared exponential kernel for its ease of understanding and its prevalence in applied spatial statistics.
This yields the outcomes model:
\begin{equation}
    \begin{split}
        & Y_{i\treat} = \underbrace{m_\treat{} + f_\treat{}(\svec_i)}_{g_\treat{}(\svec_i)} + \epsilon_i \quad\text{and}\quad
        Y_{i\ctrol} = \underbrace{m_\ctrol{} + f_\ctrol{}(\svec_i)}_{g_\ctrol{}(\svec_i)} + \epsilon_i \,; \\
        & f_\treat{}, f_\ctrol{} \overset{\indep}{\sim} \GP\del{0, k(\svec, \svec')} \quad\text{with}\quad
        k(\svec,\svec') = \sigmaf^2 \exp\del{ - \frac{\del{\svec-\svec'}\trans\del{\svec-\svec'}}{2 \ell^2}} \,.
    \end{split}
    \label{eq:spec2gp}
\end{equation}
The treatment effect at a location \(\sentinel\) on the border is derived as the difference between the two noise-free surfaces \(g_\treat{}\) and \(g_\ctrol{}\):
\begin{equation}
    \tau(\sentinel) = \sbr{m_\treat{} + f_\treat{}(\sentinel)} - \sbr{m_\ctrol{} + f_\ctrol{}(\sentinel)}\,.
\end{equation}
This can be visualized as the height of a cliff along the border \(\border\) separating the two smooth plains of the treatment and control regions.

In this specification, the hyperparameters \(\ell\), \(\sigmaf\), and \(\sigman\) are the same in the treatment and control regions, so we assume that the spatial smoothness of the responses is not affected by the treatment.
We expect that this assumption will be reasonable in many applications, but it can be easily relaxed, as discussed in \cite{Branson:2017qy}.

\subsection{Inference of the Treatment Effect}
\label{sec:inference}
If \(m_\treat\) and \(m_\ctrol\) are given normal priors with variance \(\sigmamu^2\), then the model specification \autoref{eq:spec2gp} can be used to obtain covariances between the observations, the Gaussian processes, and the mean parameters.
Given hyperparameters \(\hyperparam = \del{\ell,\sigmaf, \sigman, \sigmamu}\), any vector with entries consisting of observations, points on the potential outcomes surface \(f_{\treat}\) and \(f_{\ctrol}\), and the mean parameters \(m_{\ctrol}, m_{\treat}\) is jointly multivariate normal. Therefore the distribution of any such vector conditioned on another is also multivariate normal, with mean and covariances analytically tractable, and easily computed.

In accordance with the framework laid out in \autoref{sec:framework}, we proceed by extrapolating both Gaussian processes to the border,
and then taking the difference of the predictions to obtain the posterior treatment effect along the border.
Computationally, we need to represent this border as a set \(\sentinels=\sentinelset\) of \(\numsent\) ``sentinel'' units dotted along \(\border\).
The extrapolation step then follows mechanically through multivariate normal theory.
On the treatment side, for example:
\begin{equation}\begin{split}
    & g_\treat{}(\sentinels) \mid \yt{}, \hyperparam \sim \normal\del{\muvec_{\sentinels \mid T}, \Sigma_{\sentinels \mid T}} \,\text{, with} \\
    & \muvec_{\sentinels \mid T} =
    \KBT
    \STT^{-1} 
    \yt{} 
    \quad\text{and}\quad
    \Sigma_{\sentinels \mid T} =
    \KBB - \KBT \STT^{-1} \KBT\trans \,.
\end{split}
\label{eq:postvar2gp_t_or_c}
\end{equation}
with all the covariance matrices derived from the model specification (see \autoref{sec:covariances}).
Analogously, predictions for \(g_\ctrol{}(\sentinels)\) are obtained using the data in the control region,
and their posterior mean and covariance denoted respectively by \(\muvec_{\sentinels \mid C}\) and \(\Sigma_{\sentinels \mid C}\).
Since the two surfaces are modeled as independent, the treatment effect \(\tau(\sentinels)=g_\treat{}(\sentinels)-g_\ctrol{}(\sentinels)\) has posterior
\begin{equation}
    \begin{split}
        & \tau(\sentinels) \mid \Yvec, \hyperparam \sim \normal\del{\muvec_{\sentinels \mid Y}, \Sigma_{\sentinels \mid Y}} \,\text{, with}\\
        & \muvec_{\sentinels \mid Y} = \muvec_{\sentinels \mid \treat} - \muvec_{\sentinels \mid \ctrol} \quad\text{and}\quad
        \Sigma_{\sentinels \mid Y} = \Sigma_{\sentinels \mid \treat} + \Sigma_{\sentinels \mid \ctrol} \,.
    \end{split}
    \label{eq:postvar2gp}
\end{equation}
The posterior mean and covariance of the \(\tau(\sentinels)\) are the primary output of our GeoRDD analysis, and we refer to \autoref{eq:postvar2gp} as the ``cliff height'' estimator.

This leaves the choice of the hyperparameters: \(\hyperparam=\ell\), \(\sigmaf\), \(\sigman\), and \(\sigmamu\).
For \(\sigmamu\), we arbitrarily pick a large number, so that the prior on the mean parameters is weak.
The rest are optimized by maximizing the marginal likelihood of the observations \(\Pr\del{\Yvec \mid \ell, \sigmaf, \sigman}\), which is available analytically and easily computed for GPR.
This empirical Bayes approach is common in spatial and machine learning applications of Gaussian processes.
An alternative would be to also specify a prior on the hyperparameters, which would be preferable in order to fully account for the uncertainty in the model, but fully Bayesian inference of large Gaussian process models tends to be computationally expensive.

\subsection{Handling Nonspatial Covariates}
\label{sec:covariates}

The Gaussian process specification also makes it easy, mathematically and computationally, to incorporate a linear model on non-spatial covariates.
The models are modified by the addition of the linear regression term \(\Dmat \betavec\) on the \(n \times p\) matrix of covariates \(\Dmat\), where \(p\) is the number of non-spatial covariates.
We recommend placing a normal prior \(\normal(0,\sigmabeta^2)\) on the regression coefficients, as this preserves the multivariate normality of the model, with the simple addition of a term \(\sigma_\beta^2 \Dmat \Dmat\trans\) to the covariance \(\SigmaMat_Y\) of \(\Yvec\).
Let \(\dvec_i\) be the \(p\)-vector of non-spatial covariates of unit \(i\).
Our model becomes:
\begin{equation}
    \begin{split}
        & Y_{i\treat} = \underbrace{m_\treat{} + f_\treat{}(\svec_i)}_{g_\treat{}(\svec_i)} + \dvec_i\trans \betavec + \epsilon_i \quad\text{and}\quad
        Y_{i\ctrol} = \underbrace{m_\ctrol{} + f_\ctrol{}(\svec_i)}_{g_\ctrol{}(\svec_i)} + \dvec_i\trans \betavec + \epsilon_i \,\text{, with} \\
        & \beta_j \overset{\indep}{\sim} \normal\del{0,\sigmabeta^2}\,\text{, for }j=1,2,\dotsc,p \,,
    \end{split}
    \label{eq:covariates_model}
\end{equation}
and \(f_\treat\) and \(f_\ctrol\) as in \autoref{eq:spec2gp}.

Unfortunately, the linear term induces a covariance between the treatment and control region; \(\SigmaMat_Y\) is no longer black diagonal, which roughly quadruples the computational cost of the analysis, as it requires the inversion of an \((n_\treat{}+n_\ctrol{}) \times (n_\treat{}+n_\ctrol{})\) covariance matrix instead of (in the absence of correlations between the treatment and control units) the separate inversions of the \(n_\treat{} \times n_\treat{}\) covariance matrix \(\STT\), and \(n_\ctrol{} \times n_\ctrol{}\) covariance matrix \(\SCC\).
The introduction of the linear term modifies the cliff height estimator \autoref{eq:postvar2gp} so that its posterior mean and covariance become:
\begin{equation}
    \begin{split}
        \muvec_{\sentinels \mid Y, D} &= 
        \begingroup\SmallColSep
        \begin{bmatrix}
            \KBT & -\KBC
        \end{bmatrix}
        \SigmaMat_Y^{-1}
        \Yvec
        \endgroup
        \,\text{, and}\\
        \Sigma_{\sentinels \mid Y, D} &=
        \begingroup\SmallColSep
        2 \KBB -
        \begin{bmatrix}
            \KBT & -\KBC
        \end{bmatrix}
        \SigmaMat_Y^{-1}
        \begin{bmatrix}
            \KBT & -\KBC
        \end{bmatrix}\trans
        \endgroup
        \,.
    \end{split}
    \label{eq:cliff_with_covariates}
\end{equation}
To avoid the complexity caused by the correlation between \(\yt\) and \(\yc\) that the linear term induces, we suggest first obtaining an estimate \(\hat\betavec\) of the coefficients.
We show how to obtain the posterior mean of \(\betavec\) in \autoref{sec:betahat}.
We can then proceed with the GeoRDD analysis on the residuals, which are decorrelated conditionally on \(\betavec=\hat\betavec\).
This is an approximation, as it ignores the uncertainty in the estimate of \(\betavec\), but if the number of samples in the treatment and control areas is high (even away from the border), the approximation has negligible effect on the estimate of the treatment effect, and simplifies the subsequent GeoRDD analysis.

\section{Estimating the Local Average Treatment Effect}
\label{sec:ate}

Once we have obtained the posterior of \(\tau\) \eqref{eq:postvar2gp}, estimating the local average treatment effect (LATE) along the border will often be of interest.
We consider the class of weighted means of the functional treatment effect \(\tau\del{\sentinel}\),
with weight function \(\weightb(\sentinel)\) defined everywhere on the border \(\border\).
The weighted mean integral can be approximated as a weighted sum at the sentinels \(\sentinels\):
\begin{equation}
    \tauw = \frac{\oint_\border \left. \weightb(\sentinel) \tau(\sentinel) \dif\sentinel \right.}
    {\oint_\border \left. \weightb(\sentinel) \dif\sentinel \right.}
    \approx \frac{\sum_{\isent=1}^\numsent \weightb(\sentinel_\isent) \tau(\sentinel_\isent)}
    {\sum_{\isent=1}^\numsent \weightb(\sentinel_\isent) } \,.
\label{eq:weighted_estimand}
\end{equation}
Note that the approximation assumes that the sentinels are evenly spaced, otherwise each term in the sum needs to be re-weighted by the length of the border that the sentinel occupies.
We have shown the posterior distribution of \(\tau(\sentinels)\) to be multivariate normal, with mean \(\muvec_{\sentinels \mid Y}\) and covariance \(\Sigma_{\sentinels \mid Y}\) given in \autoref{eq:postvar2gp}.
Since \(\tauw\) is a linear transformation of \(\tau(\sentinels)\), its posterior is also multivariate normal, with mean \(\mu_{\tauw \mid Y}\) and covariance \(\Sigma_{\tauw \mid Y}\) given by
\begin{equation}
    \mu_{\tauw \mid Y} = \frac{\weightb(\sentinels)\trans \muvec_{\sentinels \mid Y}}
    {\weightb(\sentinels)\trans  \ones_\numsent}\quad\text{and}\quad
    \Sigma_{\tauw \mid Y} = \frac{\weightb(\sentinels)\trans \Sigma_{\sentinels \mid Y} \weightb(\sentinels)}
    { \del{\weightb(\sentinels)\trans  \ones_\numsent }^2 } \,,
\label{eq:weighted_estimator}
\end{equation}
where \(\weightb(\sentinels)\) is the \(\numsent\)-vector of weights evaluated at the sentinels, and \(\ones_\numsent\) is an \(\numsent\)-vector of ones.
For each estimator obtained in \autoref{eq:weighted_estimator} as a weighted mean of \(\muvec_{\sentinels \mid Y}\), we consider the ``natural'' estimand to be the same weighted mean applied to the true \(\tau\), given by \autoref{eq:weighted_estimand}.

An alternative perspective on these estimators is given by the weights induced on the observations.
Indeed, combining equations \autoref{eq:postvar2gp_t_or_c}, \autoref{eq:postvar2gp}, and \autoref{eq:weighted_estimator}, we obtain that the posterior mean of \(\tauw\) is a linear combination
\begin{equation}
    \E\del{\tauw \mid \Yvec} = \wt\trans \yt + \wc\trans \yc
    \label{eq:unit_weights}
\end{equation}
of the observed data, with ``unit weights'' given by
\begin{equation}
        \wt = \frac{
            \STT^{-1} 
            \KBT\trans \weightb(\sentinels)
        }{
            \weightb(\sentinels)\trans \ones_{\numsent}
        }
        \quad\text{and}\quad
        \wc = \frac{
            -
            \SCC^{-1} 
            \KBC\trans \weightb(\sentinels)
        }{
            \weightb(\sentinels)\trans \ones_{\numsent}
        }
        \,,
    \label{eq:unit_weights_gp}
\end{equation}
for treatment and control units respectively.

The question remains: what is the most appropriate choice of weights?
We next motivate and consider four possible choices of \(\weightb(\sentinel)\), and explore interpretations, advantages, and drawbacks. 
In \autorefexternal{sec:additional_late} of the Supplementary Materials, we discuss two further choices, the projected land LATE \(\taugeo\), and the projected superpopulation LATE \(\taupop\).
We also provide a simulation study to better understand the characteristics of the different LATE choices.
A summary of their properties is provided in \autoref{table:estimator_properties}.

\subsection{Uniform LATE}
The simplest choice is uniform weights \(\weightb(\sentinel)=1\), a seemingly reasonable and unopinionated decision.
The uniformly weighted LATE \(\unifavg\) is estimated by averaging the entries of the mean posterior at the sentinels.
Following \autoref{eq:weighted_estimand} and \autoref{eq:weighted_estimator}:
\begin{equation}\begin{split}
    &\unifavg = \oint_\border \tau(\sentinel) \dif\sentinel
        \ \big/ \ 
        \oint_\border \dif{\sentinel}  \,, \\
    &\unifavg \mid \Yvec, \hyperparam \sim \normal\del{\mu_{\unifavg \mid Y}, \Sigma_{\unifavg \mid Y}}\,\text{, with} \\
    &\mu_{\unifavg \mid Y} = \del{\ones_{\numsent}\trans \muvec_{\sentinels \mid Y}} / \numsent \quad\text{and}\quad
    \Sigma_{\unifavg \mid Y} = \del{\ones_{\numsent}\trans \Sigma_{\sentinels \mid Y} \ones_{\numsent}} / \numsent^2 \,.
\end{split}
\label{eq:unifavg}
\end{equation}
The uniformly weighted estimand takes on a geometric interpretation: equal-length segments of the border are given equal weight.
Unfortunately, uniform weights suffer from several issues that we describe and address in \autoref{sec:tau_rho} and \autoref{sec:invvar}.

\subsection{Density Weighted LATE}
\label{sec:tau_rho}

With uniform border weights, parts of the border adjoining dense populations are given equal weights to those in sparsely populated areas.
But if the border goes through an unpopulated area, such as a lake or a public park, then the treatment effect there has little meaning and importance.
Furthermore, \(\tau(\sentinel)\) in those empty areas will have large posterior variances, which will dominate the posterior variance of \(\unifavg\), potentially jeopardizing the successful detection of otherwise strong treatment effects.

We can address this issue by weighting the treatment effect at each sentinel location by the local population density \(\rho\),
i.e.\ choosing \(\weightb(\sentinel) = \rho(\sentinel)\).
Attractively, the estimand is interpretable as the average treatment effect for the superpopulation of units that live on the border:
\begin{equation}
    \taurho = \E\sbr{Y_{i\treat} - Y_{i\ctrol} \mid \svec_i \in \border}\,.
\end{equation}
It therefore better captures the ``typical'' treatment effect received by a unit than the uniformly weighted estimand.
This is the estimand used by \cite{keele_titiunik_2015}, who themselves follow in the footsteps of \cite{imbens2011regression}.

In practice, the local density needs to be estimated.
A simple kernel density estimator can be used,
though one could also deploy a more sophisticated spatial point process model.
Strictly speaking, the uncertainty of the local density estimate should then be propagated to the estimate of \(\taurho\), which may therefore no longer have a normally distributed or analytically tractable posterior.

These inconveniences certainly reduce the appeal of the density-weighted estimator,
but there is a deeper issue affecting this choice of estimand: its susceptibility to the topology of the border.
If a section of the border has more twists and turns---for example if it follows the course of a meandering river---then that section will receive disproportionately more sentinels.
The unweighted and density-weighted mean treatment estimands are both affected by this effect,
which gives higher weight to wigglier sections of the border.
See \autorefexternal{sec:wiggly_border} of the Supplementary Materials for a simulation demonstrating this susceptibility to border topology.
Consider, for illustration purposes, the border separating the two American states of Louisian and Mississippi, depicted in \autorefexternal{fig:mississippi_counties} of the Supplementary Materials.
From North to South, it follows the meandering Mississippi river, then takes a sharp turn to the East and becomes a straight line, until it meets the even more sinuous Pearl river, which it follows until it reaches the Gulf of Mexico.
Consequently, sentinels placed at equal distance intervals along this border will be more densely packed along the rivers, and sparsest along the straight segment.
When averaging a function over the border, those sections become overrepresented.
Troublingly, the sinuousness of the border therefore determines the estimand, even though the outcomes of interest will generally have nothing to do with river topologies.

\subsection{Inverse-variance Weighted LATE}
\label{sec:invvar}


This unwelcome dependence of the \(\unifavg\) and \(\taurho\) estimands on the border topology is a symptom of the geometry of the GeoRDD: the border treatment effect function \autoref{eq:functional_estimand} is defined on a 1-dimensional manifold \(\border\), which itself is embedded in a Euclidean 2-dimensional space.
The dependencies induced by this geometry are reflected in the covariance \(\Sigma_{\sentinels \mid Y}\): neighboring sentinels on a straight segment of the border will be less strongly correlated with each other than those on a sinuous segment.
The more correlated sentinels individually carry less information about the local treatment effect.
Instead of averaging the posterior treatment effect along the border based on geometry or population, we consider averaging the information contained therein.
This motivates the inverse-variance weighted mean \(\invvar\):
\begin{equation}
    \begin{split}
        & \left. \invvar \mid \Yvec, \hyperparam \right. \sim \normal\del{\mu_{\invvar \mid Y}, \Sigma_{\invvar \mid Y}} \,\text{, with} \\
        & \mu_{\invvar \mid Y} = \del{\ones_{\numsent}\trans \Sigma_{\sentinels \mid Y}^{-1} \muvec_{\sentinels \mid Y}} \big/ \del{\ones_{\numsent}\trans \Sigma_{\sentinels \mid Y}^{-1} \ones_{\numsent}} \quad\text{and}\quad
        \Sigma_{\invvar \mid Y} = 1 \big/ \del{\ones_{\numsent}\trans \Sigma_{\sentinels \mid Y}^{-1} \ones_{\numsent}} \,.
    \end{split}
    \label{eq:invvar}
\end{equation}
This estimator efficiently extracts the information from the posterior treatment effect, as it can be shown to minimize the posterior variance amongst weighted averages of the form \autoref{eq:weighted_estimand}.
It automatically gives more weight to sentinels in dense areas (as the variance will be lower there), and to sentinels in straight sections of the border (as the correlations between sentinels will be lower).

The estimand is still a weighted mean, with weights for the sentinels given by \(\weightb(\sentinels) = \Sigma_{\sentinels \mid Y}^{-1} \ones_{\numsent}\).
This can put negative weights on some sentinels, and this estimand does not lend itself to an intuitive interpretation.
It is not chosen on scientific grounds, but rather dictated by the observed data.
This is counter to the conventional approach in causal inference, that the estimand should be chosen based on substantive grounds, ideally before collecting any data.
While perhaps unorthodox, analogous ``estimands of convenience'' have been proposed in other settings, for example matching methods that exclude some unmatched units from the analysis \citep[discussed in][]{crump2009dealing}, or in the context of balancing treatment and control populations with little overlap in their covariate distributions \citep{li2016balancing}.
The 1D RDD could be said to provide another example, as the estimand \autoref{eq:rdd_univ_estimand} focuses on the treatment effect near the threshold not because those units are of particular substantive interest, but because the available data restricts estimation of the treatment effect elsewhere.

\subsection{Projected Finite-Population LATE}
All LATE estimators considered so far presuppose evenly spaced sentinel points, which are then given weights.
Alternatively, we can project the positions of treatment and control units that are within a distance \(\buffer\) of the border onto the border, and use those projected unit locations without weights (see \autorefexternal{fig:mississippi_projection_methods} of the Supplementary Materials for an illustration).
For any point \(\svec\), we use the notation \(\proj_{\border}\del{\svec}\) to give the coordinates of the point on the border \(\border\) that is closest to \(\svec\) (assuming uniqueness), and \(\dist_{\border}\del{\svec}\) for the distance between the point and the border.
Let \(\vicinity{\svec} = \allowbreak \Ind\cbr{\dist_{\border}\del{\svec} \le \buffer} \) indicate inclusion in the border vicinity.
The projected finite-population \(\tauproj\) is then the uniformly weighted mean applied with the projected unit locations instead of the evenly spaced sentinels.
We can therefore modify \autoref{eq:unifavg}, replacing the cliff height mean vector \(\muvec_{\sentinels \mid Y}\)
and covariance matrix \(\Sigma_{\sentinels \mid Y}\)
with their equivalents obtained at the projected unit locations,
to obtain the posterior mean and covariance of \(\tauproj\):
\begin{equation}\begin{split}
    &\tauproj \mid \Yvec, \hyperparam \sim \normal\del{\mu_{\tauproj \mid Y}, \Sigma_{\tauproj \mid Y}}\,\text{, with} \\
    &\mu_{\tauproj \mid Y} = 
    \sum_{i=1}^{n}
    \vicinity{\svec_i}
    \E\sbr{
        \tau\del{
            \proj_{\border}\del{\svec_i}
        }
        \mid \Yvec, \hyperparam
    } 
    \ 
    \Big/
    \ 
        \sum_{i=1}^{n}
        \vicinity{\svec_i}
    \,\text{, and}\\
    &\Sigma_{\tauproj \mid Y} = 
    \frac{
        \sum_{i=1}^{n} 
        \sum_{j=1}^{n} 
        \vicinity{\svec_i}
        \vicinity{\svec_j}
        \cov\sbr{
            \tau\del{
                \proj_{\border}\del{\svec_i}
            },
            \tau\del{
                \proj_{\border}\del{\svec_j}
            }
            \mid \Yvec, \hyperparam
        }
    }{
        \del{
            \sum_{i=1}^{n}
            \vicinity{\svec_i}
        }^2
    }
    \,.
\end{split}
\label{eq:tauproj}
\end{equation}
The posterior expectations and covariances in \autoref{eq:tauproj} are easily derived and computed analogously to the procedure of \autoref{sec:inference}.
Note that \(\tauproj\) is in the class of weighted mean estimands \autoref{eq:weighted_estimand},
with weight function \(\weightb(\sentinel) = \sum_{i=1}^{n} \vicinity{\svec_i} \delta\del{ \sentinel - \proj_\border(\svec_i)}\), where \(\delta\) is the Dirac delta function.

The resulting estimator has desirable properties: densely populated regions receive proportionately more projected units, but wigglier segments of the border do not.
While it lacks the information efficiency of the inverse-variance estimator,
the projected estimand is easier to understand and interpret,
and may feel more familiar to practitioners used to finite-population inference.
The averaging is over the observed units in the vicinity of the border, after they have been moved to the nearest point on the border.

In our experience, the choice of \(\buffer\) does not have a large effect on the estimate yielded by \autoref{eq:tauproj}.
A reasonable heuristic is to set \(\buffer\) to a small multiple of the Gaussian process lengthscale \(\ell\).
It should be noted that this choice only affects the location and density of projected units on the border; the \(\tauproj\) estimator assigns non-zero unit weights \autoref{eq:unit_weights} to all units, whether or not they fall within \(\buffer\) of the border.

\subsection{Summary}
\label{sec:summary}

The properties of the four LATE definitions proposed in this paper, and two additional choices presented in the Supplementary Materials, are summarized in \autoref{table:estimator_properties}.
In most applications, we recommend the use of the finite population or inverse-variance-weighted estimators, to prevent the undesirable influence of border topology.
The projected finite population method is simplest to understand and interpret in the tradition of finite population estimators, and unlike the density weighted LATE \(\taurho\) it does not require estimating population density.
Meanwhile, the inverse-variance estimator is the most efficient (lowest posterior variance) weighted mean estimator,
and sidesteps the choice of a distance cutoff for projected units.

\begin{table}[tbp]
    \centering
    \bgroup
    \def\arraystretch{1.1}
    \begin{tabular}{llllll}
        \hline
        Notation   & Description           & \(\border\) Topology & Sentinels & Principle & Variance \\
        \hline
        \(\unifavg\) & Uniform               & Sensitive & Equispaced      & Geometry    & High     \\
        \(\taurho\)  & Density-weighted      & Sensitive & Equispaced      & Population  & Low      \\
        \(\invvar\)  & Inverse-var. weighted & Robust    & Equispaced      & Information & Lowest   \\
        \(\tauproj\) & Projected finite pop. & Robust    & Projected       & Finite pop. & Low      \\
        \(\taugeo\)  & Proj. land            & Robust    & Proj. Grid  & Geography   & High     \\
        \(\taupop\)  & Proj. superpop.       & Robust    & Proj. Grid  & Population  & Low \\
        \hline
    \end{tabular}
    \egroup
    \caption{
    \label{table:estimator_properties}
    Summary of local average treatment effect estimator and estimand properties.}
\end{table}

\section{Testing for Non-Zero Effect}
\label{sec:hypothesis_testing}
Once we have obtained the ``cliff height'' estimate \autoref{eq:postvar2gp} and estimated a LATE, we might also naturally wonder whether we can claim to have detected a significant treatment effect at the border.
In the hypothesis testing framework, we have two possible choices of null hypotheses.
The sharp null specifies that the treatment effect is zero everywhere along the border:
\(\tau(\sentinel)=0\) for all \(\sentinel \in \border\).
Meanwhile, the weak null only requires the LATE to be zero.
We focus on a test of the weak null hypothesis here, but also provide two tests of the sharp null hypothesis based on the marginal liklihood and a chi-squared statistic in \autorefexternal{sec:alternate_tests} of the Supplementary Materials.
We found through simulations and in our applied example that the test presented in this paper has superior power and robustness to model misspecification, and therefore recommend its use.

As we saw in \autoref{sec:ate}, the LATE estimand can be defined in multiple ways.
If we choose the inverse-variance weighted mean, then \(\invvar\) has posterior given by \autoref{eq:invvar}.
While the posterior is a Bayesian object, we can use it heuristically to derive a pseudo-\(p\)-value
\(
	\tilde{p}^{\mathrm{INV}} 
        = 2\Phi\del{-
            \abs{
                \mu_{
                    \invvar \mid Y
                }
            }
            \big/
            \sqrt{
                \Sigma_{\invvar \mid Y}
            }
    }
\).
However, this pseudo-\(p\)-value obtained from the Bayesian posterior may not have good frequentist properties.
In particular, there is no guarantee that under the null hypothesis, \(p^{\mathrm{INV}}\) is below 0.05 less than 5\% of the time.

To turn it into a valid frequentist test, it can be calibrated using a parametric bootstrap under the null.
We specify a parametric null model \(\modnull\)
as a single Gaussian process spanning the control and treatment regions,
with the same kernel and hyperparameters values obtained through the procedure of \autoref{sec:inference}.
\(\modnull\) is smooth and continuous at the border,
and therefore accords with both the sharp and weak null hypotheses.
We now choose the posterior mean of the inverse-variance LATE \(\mu_{\invvar \mid Y}\) as a test statistic.
For \(b=1,\dotsc,B\) iterations, we draw \(\Yvec^{(b)}\) from \(\modnull\),
using the same spatial locations as the original data,
and compute \(\mu_{\invvar \mid Y^{(b)}}\) according to \autoref{eq:invvar} applied to the simulated data rather than the true data.
The proportion of \(\mu_{\invvar \mid Y^{(b)}}\) with absolute value greater than the observed \(\mu_{\invvar \mid Y^{obs}}\) estimates the \(p\)-value:
\begin{equation}
    p^{\mathrm{INV}} = \Pr\del{ \abs{\mu_{\invvar \mid Y}} \ge \abs{\mu_{\invvar \mid Y^{obs}}} \mid \modnull} 
    \approx \frac{1}{B} 
    \sum_{b=1}^B 
        \Ind\cbr{
            \abs{
                \mu_{\invvar \mid Y^{(b)}}
            } 
            \ge  
            \abs{
                \mu_{\invvar \mid Y^{obs}}
            } 
        }
    \,.
\end{equation}
Computationally, because the hyperparameters and locations of the units are held constant during the bootstrap, we can reuse the Cholesky decomposition of the covariance matrix, allowing the test to be performed in seconds even with hundreds of units and thousands of bootstrap samples.

The calibration can also be achieved analytically, since \(\mu_{\invvar \mid Y}\) is normally distributed under the null hypothesis.
We derive the analytical calibration of hypothesis tests based on any LATE estimand in \autoref{sec:calibration}.
Note that the \(p\)-value for this test is derived under the parametric null model \(\modnull\), which accords with both the sharp null and weak null hypotheses, but is not the only possible model that satisfies the weak null.
The calibrated inverse-variance test “targets” the weak null hypothesis in the sense that the test statistic is an estimate of the LATE, and thus the test is sensitive to deviations of the LATE from zero, rather than its \(p\)-value being derived directly under the weak null (such as the classical \(t\)-test).

\label{eq:calib_test}

\subsection{Placebo Tests}
\label{sec:placebo}
Gaussian process models are almost always misspecified.
We do not believe that the Gaussian process with stationary squared exponential kernel is the true data-generating process, although we hope that the model is sufficiently flexible to represent reality well.
Under misspecification, we should be skeptical of results that rely on the truth of the model specification.
We therefore encourage practitioners to probe the validity of the hypothesis test by running a ``placebo'' test.
A placebo test repeatedly applies the hypothesis test on data that are known to have zero treatment effect (a ``placebo''),
in order to verify that the returned \(p\)-values are uniformly distributed.
In our spatial setting, we use the treatment and control regions separately as placebo groups.
Within each placebo group, we repeatedly draw an arbitrary geographical border, creating new treatment and control groups.
Here we drew lines that split the placebo units in half at a sequence of angles \(1\degree,2\degree,3\degree,\dotsc,180\degree\) counter-clockwise from horizontal, each positioned so that half of the units fall on either side of the line in order to maximize power.
Because the border was chosen arbitrarily by us, without reference to the outcomes, we should not expect to see a discontinuous jump in outcomes at this border.
We apply the calibrated inverse-variance test procedure described above to this arbitrarily divided data, store the results, and hope to obtain a roughly uniform distribution of \(p\)-values.
The resulting \(p\)-values will obviously be highly correlated, so we should only expect a very roughly uniform distribution (because of the small effective sample size), but at the very least, this procedure allows us to visually verify that the \(p\)-values are not blatantly biased.

\section{Application: NYC School Districts}
\label{sec:NYC_example}

We illustrate the analysis of a GeoRDD with house sales data in New York City.
The city publishes information pertaining to property sales within the city in the last 12 months on a rolling basis,
available at \url{https://www1.nyc.gov/site/finance/taxes/property-rolling-sales-data.page},
which we downloaded on September~15, 2016.
The dataset includes columns for the sale price, building class, and the address of the property.
Public schools in the city are all part of the City School District of the City of New York, but the city-wide district is itself divided into 32 sub-districts.
It is a common belief that school districts have an impact on real estate price, as parents are willing to pay more to live in districts with better schools.
We therefore ask: can we measure a discontinuous jump in house prices across the borders separating school districts?

\begin{figure}[tb]
    \centering
    \includegraphics[width=\textwidth,height=0.4\textheight,keepaspectratio]{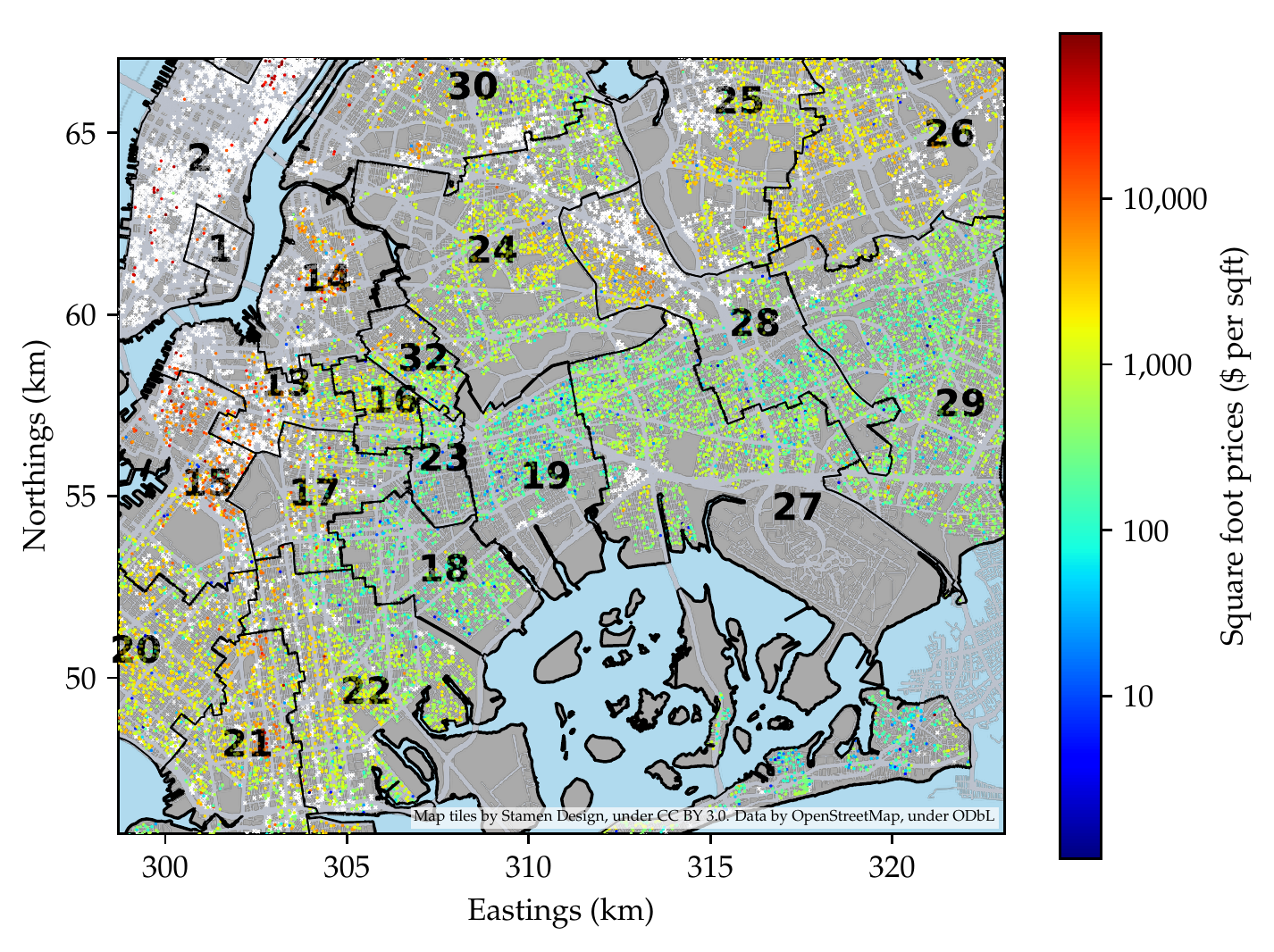}
    \caption{\label{fig:sales_map}Map of property sales in New York City. Each dot is a sale, and its color indicates the price per square foot. White crosses indicate sales of properties with missing square footage, which are therefore excluded from the analysis. School district boundaries are shown, and each district is labeled by its number.}
\end{figure}

In order to model the property sale prices, we first need to obtain their locations.
We geocode the address of each sale by merging the sales with NYC's Pluto database, which contains X and Y coordinates for each house, identified by its borough, zip code, block and lot.
These coordinates are given in the EPSG:2263 projection, which we also adopt.
For addresses that do not find a match in Pluto, we use Google's geocoding API to obtain a latitude and longitude, which we then project onto EPSG:2263.

We then filter the 56,815 sales, by removing
36,448 sales outside of the family homes building class categories (one, two, and three family dwellings),
4 remaining sales missing the square footage information,
and 785 remaining sales with outlier log price per square foot less than 3 or more than 8.
We exclude condos and coops because only very few sales report square footage alongside the price.
The resulting dataset of 19,578 sales is displayed in \autoref{fig:sales_map}.
The 27,394 residential properties with missing square footage information are also shown;
these are almost all coops and condos, which explains the clustering of missing data in areas of higher density.

\subsection{Model for Property Prices}
The outcome of interest is price per square foot.
As is commonly done in analyses of real estate prices, we take its logarithm to reduce the skew of the outcome.
The complete model is then a Gaussian process within each district (indexed by \(j=1,\dotsc,J_\district\)) over the spatial covariates \(\svec\), super-imposed with a linear regression on the property covariates (which are \(L_\building\) building categories encoded as dummy variables).
This model can be written:
\begin{equation}
    \begin{aligned}
        Y_i &= 
        	m_{\district\sbr{i}} + \beta_{\building\sbr{i}}
        	+ f_{\district\sbr{i}}(\svec_i) + \epsilon_i 
			\,,\quad
 			\epsilon_i \sim \normal\del{0, \sigma_y^2} 
			\,,
			\phantom{\hspace{-10cm}} 
            \\
        \beta_{l} &\sim \normal\del{0, \sigma_\beta^2}
			\,,
			&&
			\text{ for }
			l=1,\dotsc,L_\building \,, \\
        m_{j} &\sim \normal\del{0, \sigmamu^2}
			\,,
			&&
			\text{ for }
			j=1,\dotsc,J_\district \,, \\
        f_j &\sim \GP\del{0, k(\svec, \svec')}
			\,,
			&&
			\text{ for }
			j=1,\dotsc,J_\district \,,
    \end{aligned}
    \label{eq:nyc_model}
\end{equation}
where \(k\) is the squared exponential covariance function as in \autoref{eq:spec2gp}.

A visual inspection of the house sales map in \autoref{fig:sales_map} drew our attention towards the border between districts 19 and 27.
We arbitrarily designate district 19 as the ``treatment'' area and district 27 as the ``control'' area.
Importantly, the border between the two districts is also part of the border between Brooklyn and Queens, so we will not be able to attribute a difference in price solely to the causal effect of the school districts.
This is an instance of what \cite{keele_titiunik_2015} term ``compound treatments,'' a frequent concern in GeoRDDs.
Therefore, we are mainly \emph{measuring} a discontinuity in the house prices at the border.
Attributing the discontinuity to a particular cause (school district or borough) is not directly supported by the data.

Another concern is units sorting around the border, which would violate the identification assumptions for GeoRDDs.
If people move across the border to live in a better school district, does this invalidate the analysis?
We take the view that the unit of analysis here is the tract of land on which houses are built, rather than the residents themselves.
If a district becomes more attractive, people may move to it, whereas land does not move but its price adjusts.
A sale gives a snapshot of the price of the land, made more accurate by correcting in our model \autoref{eq:nyc_model} for covariates that pertain to the building rather than land.
Note that of course, the limited covariates provided by the data cannot fully capture the value of the building.
For example, the wealthier residents who inhabit the more desirable school districts may also have more funds available to maintain and enhance their home, which will drive up the property's resale value.
Since it is not captured by the available covariates, this added value is folded into the treatment effect by our analysis.

\begin{figure}[tb]
    \centering
    \includegraphics[width=0.5\textwidth]{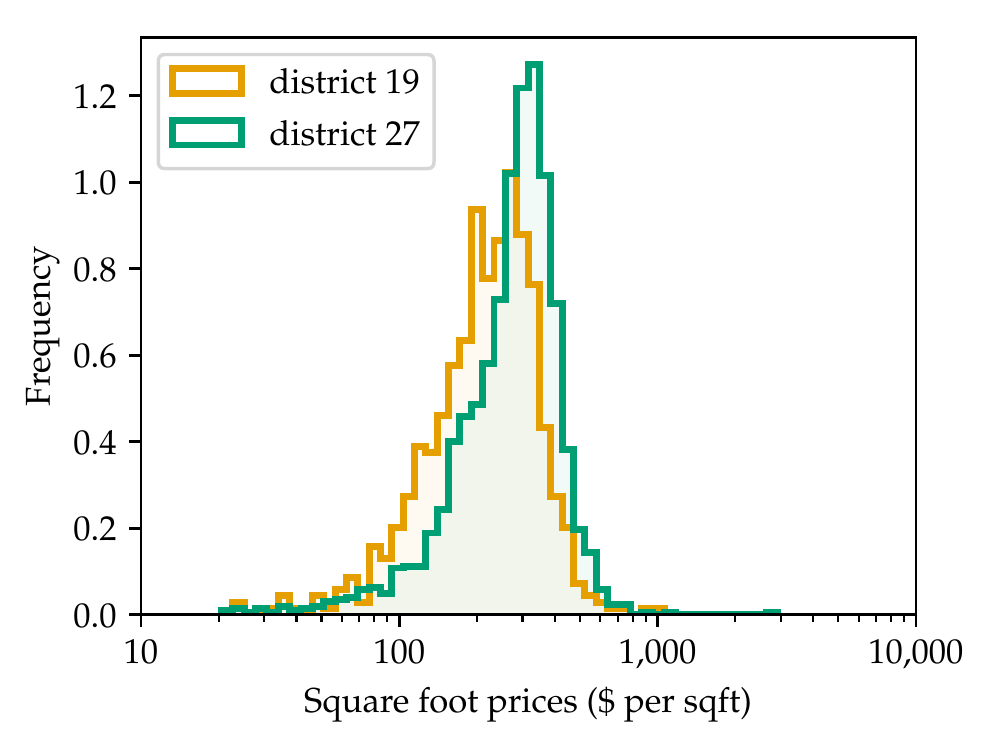}
    \caption{\label{fig:NYC_histogram}Histogram of log sale prices per square foot of houses sold in NYC school districts 19 and 27.}
\end{figure}

The histogram in \autoref{fig:NYC_histogram} of log prices per square foot for sales in both districts shows that marginally the house prices are very different.
Our goal is to establish whether this difference is measurable at the border, and not merely an underlying trend that spans both districts.

We fit the hyperparameters \(\sigma_\beta\), \(\sigmaf\), \(\ell\) and \(\sigman\) by optimizing the marginal log-likelihood of the data within neighboring school districts 18, 19, 23, 24, 25, 26, 27, 28, and 29.
We hold \(\sigmamu\) fixed to 20 to give the district means \(m_j\) a weak prior.
The fitted hyperparameters were \(\widehat{\sigman}=0.40\), \(\widehat{\sigmaf}=0.20\), \(\widehat{\sigma_\beta}=0.15\), and \(\widehat{\ell}=1.4~\text{km}\).

\subsection{Cliff Height Estimator}

We seek to estimate the treatment effect function \(\tau\) on the border between the two districts.
We could proceed by computing the cliff height estimator with covariates \autoref{eq:cliff_with_covariates}.
But to simplify the analysis as discussed in \autoref{sec:covariates}, we can instead obtain the posterior means of the \(\beta\) coefficients (following the procedure outlined in \autoref{sec:betahat}, but extended to \(J_\district\) rather than just two areas), and extract the residuals \(\Yvec{}-\Dmat \hat{\betavec}\).
We then treat the residuals as the observed outcomes in a GeoRDD analysis with no non-spatial covariates.
In this example, we find that the posterior variance of \(\betavec\) is low, and therefore the two approaches yield very similar results, but conditioning on the estimate of \(\betavec\) is computationally convenient.

\begin{figure}[tb]
    \centering
    \includegraphics[height=0.35\textheight]{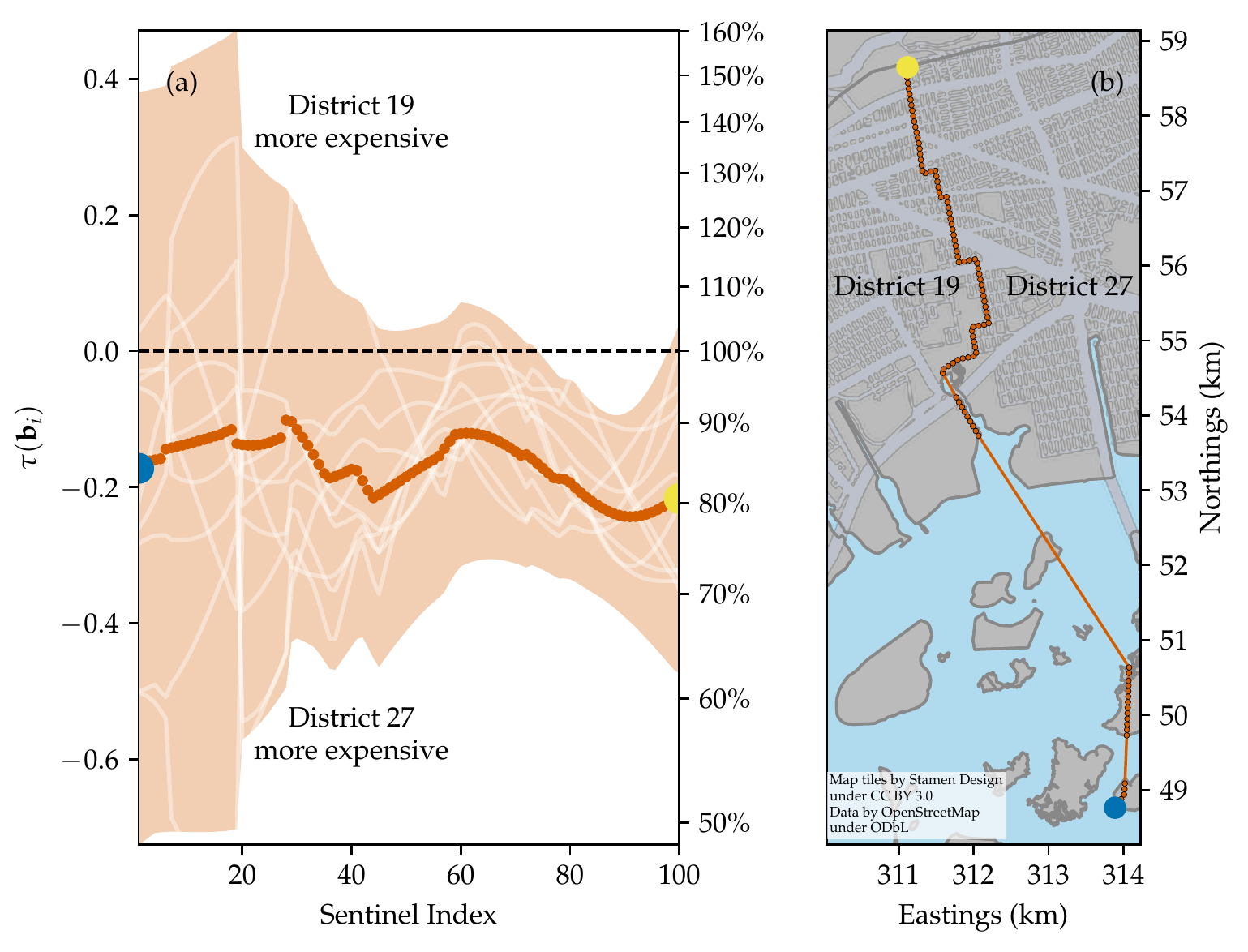}
    \caption{\label{fig:NYC_cliff_height}
        (a)
        Cliff height estimator \autoref{eq:postvar2gp} for the school district effect on house prices per square foot between district 27 and district 19, with 95\% credible envelope.
        The left \(y\)-axis shows the difference in log prices per square foot; positive values mean prices are higher in district 19.
        The right \(y\)-axis shows the corresponding ratio of the price of a house in district 19 over its price in district 27.
        A few draws from the posterior are shown in lighter color to show the posterior correlations between sentinels.
        Note the decorrelation from sentinels 6 to 7, and 19 to 20, where the border crosses the water between sparsely populated islands in Jamaica Bay and then onto Long Island.
        (b)
        A map of sentinel locations, evenly spaced along the border between school districts 27 and 19.
    	The southernmost sentinel (shown as a blue circle in both plots) has index 1, while the northernmost sentinel (shown in yellow) has index 100.
	}
\end{figure}

Following the inference procedure outlined in \autoref{sec:twogp}, we obtain the posterior distribution of the cliff height \(\tau(\sentinels)\) obtained at \(R=100\) sentinel locations evenly spaced along the border.
The cliff height is shown in \autoref{fig:NYC_cliff_height}, and shows that \(\tau\) is estimated as negative everywhere along the border, which corresponds to higher property prices in district 27.
However, the credible envelope is fairly wide, especially in the southern section of the border, so we cannot visually rule out the null hypothesis that \(\tau=0\).

\subsection{Average Log-Price Increase}
The cliff height \autoref{fig:NYC_cliff_height} shows a negative treatment effect everywhere along the border, which can be averaged by the estimators we developed in \autoref{sec:ate}.
Our two recommended estimators, based on inverse-variance weighting and finite-population projection of units within \(\buffer=2\ell\) of the border, yield LATE estimates of \(-0.19\) and \(-0.18\) respectively, which corresponds to an almost 20\% increase in property prices going from district 19 to district 27.
By contrast, treating each district and building class as a fixed effect in an ordinary least squares (OLS) model yields a treatment effect estimate (the difference between the district 19 and 27 coefficients) of -0.12. 
This smaller estimate could be explained by an overall East to West positive spatial trend in prices, visible between districts 29 and 15 in \autoref{fig:sales_map}, which would confound the OLS estimate of the treatment effect.
All LATE estimators from \autoref{sec:ate} applied to this setting are shown in \autoref{table:NYC_ate}.
In this example the different estimators yield similar answers, as the border is fairly straight and short relative to the fitted lengthscale.

\begin{table}
    \centering
    \begin{tabular}{lrrr}
        \hline
        & \multicolumn{3}{c}{Posterior} \\
        Estimand & Mean & Standard Dev. & Tail Prob. \\
        \hline
	      \(\unifavg\) & -0.17 & 0.08 & 1.98\% \\
	      \(\taurho\) & -0.19 & 0.06 & 0.04\% \\
	      \(\invvar\) & -0.19 & 0.06 & 0.03\% \\
	      \(\tauproj~(\buffer=2\ell)\) & -0.18 & 0.06 & 0.13\% \\
		  \(\taugeo~(\buffer=2\ell)\) & -0.16 & 0.09 & 3.80\% \\
		  \(\taupop~(\buffer=2\ell)\) & -0.18 & 0.06 & 0.15\% \\
        \hline
    \end{tabular}
    \caption{
    \label{table:NYC_ate}
Average difference in log price per square foot between school districts 19 and 27. For each LATE estimand, we show the mean and standard deviation of its posterior distribution, and the tail probability \(\Pr(\tau > 0 \mid \Yvec, \hat{\beta}, \hyperparam)\). 
Negative LATEs correspond to district 27 being more expensive.
}
\end{table}

\subsection{Significant Difference in Price?}
The estimated inverse-variance weighted mean treatment effect is suggestive of a significant treatment effect.
But the posterior tail probability cannot be interpreted as a \(p\)-value.
For this, we turn to the test developed in \autoref{sec:hypothesis_testing}, which yields a \(p\)-value of \(p^{\mathrm{INV}}=0.002\), thus rejecting the null hypothesis that there is no difference in house prices at the border between districts 19 and 27.

\begin{figure}[tb]
    \centering
    \includegraphics[width=\textwidth,height=0.3\textheight,keepaspectratio]{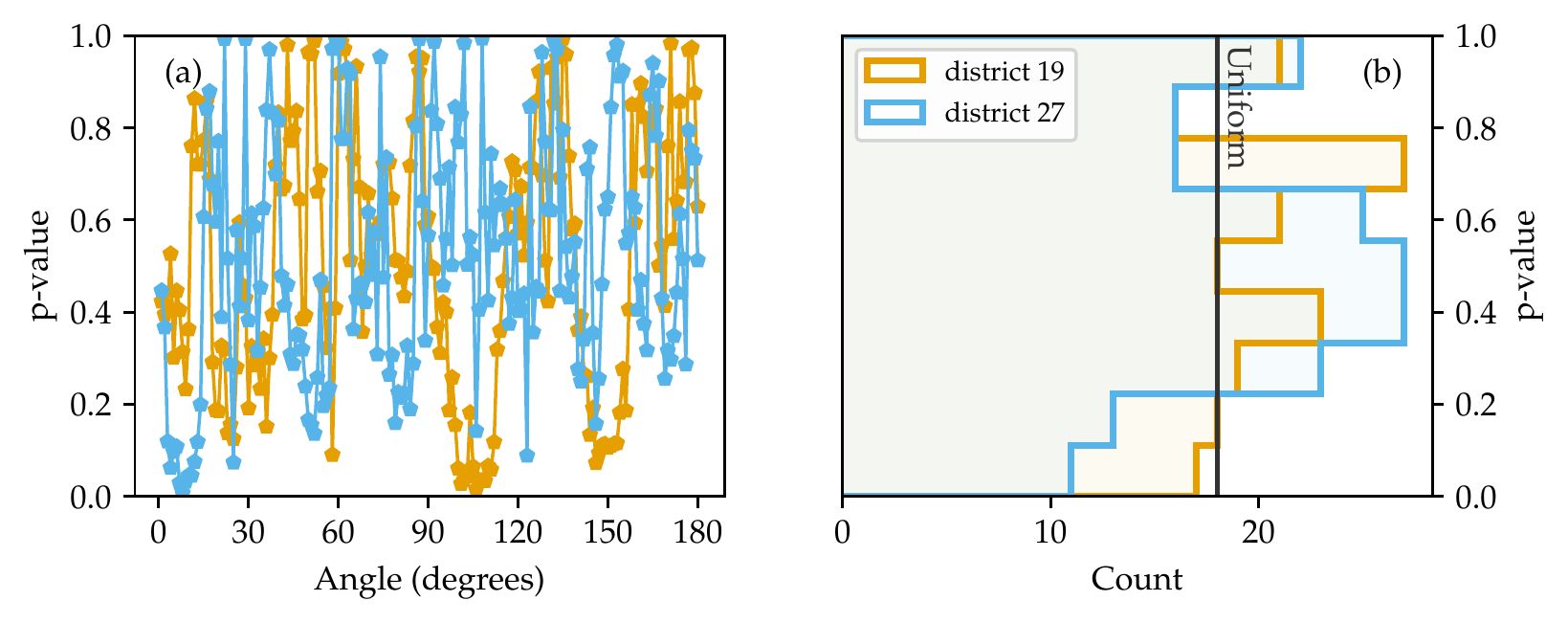}
    \caption{\label{fig:placebo_invvar} Placebo tests for calibrated inverse-variance test of difference in house prices at the border between NYC school districts 19 and 27. 
    (a) the placebo \(p\)-value as a function of the border angle;
    (b) histogram of the placebo \(p\)-values, with the black vertical line indicating the uniform distribution.}
\end{figure}

To assess the validity of the test, we apply the placebo tests devised in \autoref{sec:placebo},
the results of which are shown in \autoref{fig:placebo_invvar}.
Within each district, we split the data in half by a line at angles \(1\degree\), \(2\degree\), \(\dotsc\), \(180\degree\).
Because these lines were drawn arbitrarily, we do not expect a discontinuous treatment effect between the two halves, and so we hope to see a uniform distribution of placebo \(p\)-values.
However, these tests will be highly correlated---there is in fact noticeable autocorrelation in the graph of the placebo \(p\)-value as a function of angle---and so the low effective sample size could lead to some apparent departures from uniformity.
Nonetheless, we do not observe a flagrant bias in \autoref{fig:placebo_invvar}(b), which therefore does not discredit the calibrated inverse-variance test, and confirms the significance of the difference in price at the border between the two districts.

\section{Conclusion}

Geographic regression discontinuity designs (GeoRDDs) arise when a treatment is assigned to one region, but not to another adjacent region.
For outcomes that vary spatially, a direct comparison of mean outcomes between \(\yt\) and \(\yc\), such as a \(t\)-test, is an invalid estimator of the treatment effect, as it is confounded by the spatial covariates.
However, under smoothness assumptions, units adjacent to the border are comparable, and form a natural experiment.
The same idea underpins causal interpretations of one-dimensional regression discontinuity designs (1D RDDs), where a single ``forcing'' variable controls the treatment assignment instead of a border separating two geographical regions.
We use this similarity to motivate a framework for the analysis of GeoRDDs, which proceeds in three steps:
\begin{flatlist} 
\item fit a smooth surface on either side of the border,
\item extrapolate the surfaces to the border, and 
\item take the difference of the two extrapolations to estimate the treatment effect along the border.
\end{flatlist}

Previous research has focused on extending methods developed for 1D RDDs to GeoRDDs.
In applied settings, some have used the signed distance from the border as the forcing variable in a 1D RDD, but the resulting estimator is spatially confounded.
In this paper, we emphasize the importance of the spatial aspect of the design, and therefore draw from the spatial statistics literature, which brings a rich set of tools designed to model and exploit spatial correlations.
We used Gaussian process regression, known as kriging in the spatial statistics literature, to fit the smooth surfaces to the outcomes in step (1) of our framework.
Our approach yields a multivariate normal posterior distribution of the treatment effect for a collection of ``sentinel'' locations along the border.

Averaging the treatment effect along the border turns out to have surprising pitfalls.
Simply integrating the treatment effect uniformly along the border yields an estimand that is inefficient and undesirably sensitive to the topology of the border.
More sophisticated estimands, summarized in \autoref{table:estimator_properties}, are robust to this effect, and use the information available in the data more efficiently.

To test against the null hypothesis of zero treatment effect along the border, we develop a test based on the posterior distribution of the LATE.
We use the inverse-variance weighted LATE to attain high power, but the other LATE estimates of \autoref{sec:ate} could be used similarly.
To ensure good frequentist properties we “calibrate” the test, obtaining its distribution under the null model, either using a parametric bootstrap or analytically.

We applied our method to a publicly available dataset of one year of New York City property sales, to examine whether school district cause difference in property prices.
Focusing on the border between school districts 19 and 27, we estimated a roughly 20\% average increase in house prices per square foot when crossing the border from district 19 to district 27.
However, the border between these two districts is also the border between the NYC boroughs of Brooklyn and Queens, so we cannot attribute this difference to the causal effect of the school districts.
In the Supplementary Materials, we extend the GeoRDD analysis to other pairs of adjacent school districts, and find significant effects between many of these pairs.
However, in some of these cases, physical barriers like parks, commercial zones, railways, and major roads can separate neighborhoods, keep data away from the borders, break the stationarity assumption of the spatial model, and increase the amount of extrapolation performed by the model, which casts doubt on the legitimacy of the estimated treatment effects.
Missing data from condo sales which do not report square footage can also distort estimated effects.
Overall, it seems that school district borders in Brooklyn and Queens are often accompanied by a discontinuity in house prices, but the causal attribution of this difference to the reputation of the school districts can be questionable.

The main limitation of our approach to GeoRDDs is the reliance on modeling assumptions.
We modeled the response surfaces as two independent Gaussian processes, with iid normal noise for each observation.
As is common in spatial statistics, we use Gaussian process regression as a non-parametric smoothing device that flexibly captures spatial correlations, but do not claim that our model is a true representation of the stochastic mechanism generating the data.
We believe care must therefore be taken not to lean heavily on modeling assumptions.
In particular, we recommend that hypothesis tests always be accompanied by placebo tests:
by applying the same procedure with arbitrary borders where no treatment was applied, we can verify that the test behaves appropriately under the null hypothesis, despite any potential model misspecification.
We also assumed a stationary covariance structure, with hyperparameters equal in the treatment and control regions, and in particular we chose the squared exponential kernel.
This kernel makes smoothness assumptions that are often considered unrealistic in geostatistical settings; the Mat\'ern covariance family is often recommended as a more robust alternative.
The assumption of equal covariance parameters in the two areas can also be relaxed, by separately tuning the parameters within each area.

Because of the need to extrapolate the fitted processes a short distance to the border, our GeoRDD method may be vulnerable to the limitations of Gaussian processes when extrapolating.
The distinction between interpolation and extrapolation of spatial models is explored in some depth in \cite{stein2012interpolation}.
We expect that methodological advances that improve the extrapolating behavior of Gaussian processes would also improve the robustness of our method.
For example, \cite{wilson2013gaussian} develop spectral mixture (SM) covariance kernels with good extrapolating behavior, which could be applied beneficially to GeoRDDs.
However, SM kernels are motivated by time series with some periodic or oscillatory behavior, which is more unusual in spatial applications, and may therefore not be as well-suited for use with GeoRDDs.

The use of GPR to analyse GeoRDDs gives flexibility and extensibility to the method.
This presents many opportunities for future research, inspired by the past and future development of methods in spatial statistics and machine learning that are based on Gaussian processes.
In spatial statistics, kriging has been used as the foundation for a plethora of spatial models, which may be adapted for the purposes of analyzing GeoRDDs.
\cite{banerjee2014hierarchical} provides a good introduction to the richness of the spatial statistics field.
For example, if the outcomes are binary, proportions, or counts, then binomial or Poisson likelihoods could be substituted instead of the normal likelihood used in this paper.

Furthermore, in some applications, it may be of substantive interest to know whether the treatment effect is constant (homogenous) or variable (heterogenous).
Hypothesis tests targeting the homogeneity of the treatment effect along the border would be an interesting possible extension of our framework.

The framework and techniques of this paper could also be extended to spatio-temporal settings.
If the treatment is only applied to the treatment region after a time \(t^*\), one could envision a three-dimensional RDD consisting of the geographical border in the spatial dimensions, and a straight line through \(t^*\) in the temporal dimension.
The Gaussian process model would need to be augmented with a temporal component, for example with an anisotropic squared exponential covariance function.
We leave spatio-temporal RDDs using Gaussian process models to future research.

\bibliographystyle{chicago}
\bibliography{GeoRDD}

\begin{thebibliography}{}

\bibitem[\protect\citeauthoryear{Antonelli, Cefalu, and Bornn}{Antonelli
  et~al.}{2016}]{antonelli2016positive}
Antonelli, J., M.~Cefalu, and L.~Bornn (2016).
\newblock The positive effects of population-based preferential sampling in
  environmental epidemiology.
\newblock {\em Biostatistics\/}~{\em 17\/}(4), 764--778.

\bibitem[\protect\citeauthoryear{Banerjee, Carlin, and Gelfand}{Banerjee
  et~al.}{2014}]{banerjee2014hierarchical}
Banerjee, S., B.~P. Carlin, and A.~E. Gelfand (2014).
\newblock {\em Hierarchical modeling and analysis for spatial data}.
\newblock Crc Press.

\bibitem[\protect\citeauthoryear{Branson, Rischard, Bornn, and
  Miratrix}{Branson et~al.}{2017}]{Branson:2017qy}
Branson, Z., M.~Rischard, L.~Bornn, and L.~Miratrix (2017, 04).
\newblock A nonparametric bayesian methodology for regression discontinuity
  designs.

\bibitem[\protect\citeauthoryear{Chen, Ebenstein, Greenstone, and Li}{Chen
  et~al.}{2013}]{chen2013evidence}
Chen, Y., A.~Ebenstein, M.~Greenstone, and H.~Li (2013).
\newblock Evidence on the impact of sustained exposure to air pollution on life
  expectancy from china's huai river policy.
\newblock {\em Proceedings of the National Academy of Sciences\/}~{\em
  110\/}(32), 12936--12941.

\bibitem[\protect\citeauthoryear{Cook}{Cook}{2008}]{cook2008waiting}
Cook, T.~D. (2008).
\newblock ``waiting for life to arrive'': a history of the
  regression-discontinuity design in psychology, statistics and economics.
\newblock {\em Journal of Econometrics\/}~{\em 142\/}(2), 636--654.

\bibitem[\protect\citeauthoryear{Crump, Hotz, Imbens, and Mitnik}{Crump
  et~al.}{2009}]{crump2009dealing}
Crump, R.~K., V.~J. Hotz, G.~W. Imbens, and O.~A. Mitnik (2009).
\newblock Dealing with limited overlap in estimation of average treatment
  effects.
\newblock {\em Biometrika\/}~{\em 96\/}(1), 187--199.

\bibitem[\protect\citeauthoryear{Ding}{Ding}{2014}]{Ding:2014sf}
Ding, P. (2014, 02).
\newblock A paradox from randomization-based causal inference.

\bibitem[\protect\citeauthoryear{Hahn, Todd, and Van~der Klaauw}{Hahn
  et~al.}{2001}]{hahn2001identification}
Hahn, J., P.~Todd, and W.~Van~der Klaauw (2001).
\newblock Identification and estimation of treatment effects with a
  regression-discontinuity design.
\newblock {\em Econometrica\/}~{\em 69\/}(1), 201--209.

\bibitem[\protect\citeauthoryear{Imbens and Kalyanaraman}{Imbens and
  Kalyanaraman}{2012}]{imbens2012optimal}
Imbens, G. and K.~Kalyanaraman (2012).
\newblock Optimal bandwidth choice for the regression discontinuity estimator.
\newblock {\em The Review of economic studies\/}~{\em 79\/}(3), 933--959.

\bibitem[\protect\citeauthoryear{Imbens and Zajonc}{Imbens and
  Zajonc}{2011}]{imbens2011regression}
Imbens, G. and T.~Zajonc (2011).
\newblock Regression discontinuity design with multiple forcing variables.
\newblock {\em Report, Harvard University.[972]\/}.

\bibitem[\protect\citeauthoryear{Imbens and Lemieux}{Imbens and
  Lemieux}{2008}]{imbensrdd}
Imbens, G.~W. and T.~Lemieux (2008).
\newblock Regression discontinuity designs: A guide to practice.
\newblock {\em Journal of econometrics\/}~{\em 142\/}(2), 615--635.

\bibitem[\protect\citeauthoryear{Keele, Lorch, Passarella, Small, and
  Titiunik}{Keele et~al.}{2017}]{keeleoverview}
Keele, L., S.~Lorch, M.~Passarella, D.~Small, and R.~Titiunik (2017).
\newblock {\em An Overview of Geographically Discontinuous Treatment
  Assignments with an Application to Children's Health Insurance}, Chapter~4,
  pp.\  147--194.
\newblock Emerald Publishing Limited.

\bibitem[\protect\citeauthoryear{Keele, Titiunik, and Zubizarreta}{Keele
  et~al.}{2015}]{keele2015enhancing}
Keele, L., R.~Titiunik, and J.~R. Zubizarreta (2015).
\newblock Enhancing a geographic regression discontinuity design through
  matching to estimate the effect of ballot initiatives on voter turnout.
\newblock {\em Journal of the Royal Statistical Society: Series A (Statistics
  in Society)\/}~{\em 178\/}(1), 223--239.

\bibitem[\protect\citeauthoryear{Keele and Titiunik}{Keele and
  Titiunik}{2015}]{keele_titiunik_2015}
Keele, L.~J. and R.~Titiunik (2015).
\newblock Geographic boundaries as regression discontinuities.
\newblock {\em Political Analysis\/}~{\em 23\/}(1), 127--155.

\bibitem[\protect\citeauthoryear{Li, Morgan, and Zaslavsky}{Li
  et~al.}{2016}]{li2016balancing}
Li, F., K.~L. Morgan, and A.~M. Zaslavsky (2016).
\newblock Balancing covariates via propensity score weighting.
\newblock {\em Journal of the American Statistical Association\/}.

\bibitem[\protect\citeauthoryear{MacDonald, Klick, and Grunwald}{MacDonald
  et~al.}{2015}]{macdonald2015effect}
MacDonald, J.~M., J.~Klick, and B.~Grunwald (2015).
\newblock The effect of private police on crime: evidence from a geographic
  regression discontinuity design.
\newblock {\em Journal of the Royal Statistical Society: Series A (Statistics
  in Society)\/}.

\bibitem[\protect\citeauthoryear{Papay, Willett, and Murnane}{Papay
  et~al.}{2011}]{papay2011extending}
Papay, J.~P., J.~B. Willett, and R.~J. Murnane (2011).
\newblock Extending the regression-discontinuity approach to multiple
  assignment variables.
\newblock {\em Journal of Econometrics\/}~{\em 161\/}(2), 203--207.

\bibitem[\protect\citeauthoryear{Rasmussen and Williams}{Rasmussen and
  Williams}{2006}]{rasmussen2006gaussian}
Rasmussen, C.~E. and C.~K. Williams (2006).
\newblock {\em Gaussian processes for machine learning}, Volume~1.
\newblock MIT press Cambridge.

\bibitem[\protect\citeauthoryear{Rencher}{Rencher}{2003}]{rencher2003methods}
Rencher, A.~C. (2003).
\newblock {\em Methods of multivariate analysis}, Volume 492.
\newblock John Wiley \& Sons.

\bibitem[\protect\citeauthoryear{Stein}{Stein}{2012}]{stein2012interpolation}
Stein, M.~L. (2012).
\newblock {\em Interpolation of spatial data: some theory for kriging}.
\newblock Springer Science \& Business Media.

\bibitem[\protect\citeauthoryear{Thistlethwaite and Campbell}{Thistlethwaite
  and Campbell}{1960}]{thistlethwaite1960regression}
Thistlethwaite, D.~L. and D.~T. Campbell (1960).
\newblock Regression-discontinuity analysis: An alternative to the ex post
  facto experiment.
\newblock {\em Journal of Educational psychology\/}~{\em 51\/}(6), 309.

\bibitem[\protect\citeauthoryear{Wilson and Adams}{Wilson and
  Adams}{2013}]{wilson2013gaussian}
Wilson, A. and R.~Adams (2013).
\newblock Gaussian process kernels for pattern discovery and extrapolation.
\newblock In {\em Proceedings of the 30th International Conference on Machine
  Learning (ICML-13)}, pp.\  1067--1075.

\bibitem[\protect\citeauthoryear{Zubizarreta}{Zubizarreta}{2012}]{zubizarreta2012using}
Zubizarreta, J.~R. (2012).
\newblock Using mixed integer programming for matching in an observational
  study of kidney failure after surgery.
\newblock {\em Journal of the American Statistical Association\/}~{\em
  107\/}(500), 1360--1371.

\end{thebibliography}

\begin{appendices}

\section{Covariances for Gaussian Process Model}
\label{sec:covariances}
All covariances below are conditional on the hyperparameters \(\hyperparam = \del{\ell,\sigmaf, \sigman, \sigmamu}\), omitted for concision.
\begin{equation}
    \begin{split}
        m_\treat, m_\ctrol   &\sim \normal\del{0,\sigmamu^2} \\
        \cov(Y_{i\treat},m_\treat)  &= \cov(Y_{i\ctrol},m_\ctrol) = \sigmamu^2 \\
        \cov(Y_{i\treat},m_\ctrol)  &= \cov(Y_{i\ctrol},m_\treat)  = 0 \\
        \cov\del{Y_{i\treat}, f_{\treat}(\svec')} &= \cov\del{Y_{i\ctrol}, f_{\ctrol}(\svec')} = k(\svec_i,\svec') \\
        \cov\del{Y_{i\treat}, f_{\ctrol}(\svec')} &= \cov\del{Y_{i\ctrol}, f_{\treat}(\svec')} = 0 \\
        \cov(Y_{i\treat},Y_{j\treat}) &= \cov(Y_{i\ctrol},Y_{j\ctrol}) = \sigmamu^2 + k(\svec_i,\svec_j) + \delta_{ij}\sigman^2 \\
        \cov(Y_{i\treat},Y_{j\ctrol}) &= 0
    \end{split}
    \label{eq:covariances}
\end{equation}
We further define some shorthand notation, found in \autoref{table:notation}.

\begin{table}[bp]
    \centering
    \bgroup
    \def\arraystretch{1.2}
    \begin{tabular}{lll}
        \hline
        Symbol & Size                       & \(ij^{\mathrm{th}}\) entry                                                      \\ \hline
        \(\KBB\) & \(\numsent \times \numsent\) & \(\sigmamu^2 + k\del{\sentinel_i,\sentinel_j}\)                                 \\ 
        \(\KBT\) & \(\numsent \times n_\treat\) & \(\sigmamu^2 + k\del{\sentinel_i,\svec_{j\treat}}\)                             \\ 
        \(\KBC\) & \(\numsent \times n_\ctrol\) & \(\sigmamu^2 + k\del{\sentinel_i,\svec_{j\ctrol}}\)                             \\
        \(\KTT\) & \(n_\treat \times n_\treat\) & \(\sigmamu^2 + k\del{\svec_{i\ctrol},\svec_{j\ctrol}}\)                         \\
        \(\KCC\) & \(n_\ctrol \times n_\ctrol\) & \(\sigmamu^2 + k\del{\svec_{i\treat},\svec_{j\treat}}\)                         \\ 
        \(\STT\) & \(n_\treat \times n_\treat\) & \(\sigmamu^2 + k\del{\svec_{i\treat},\svec_{j\treat}} + \delta_{ij} \sigman^2\) \\ 
        \(\SCC\) & \(n_\ctrol \times n_\ctrol\) & \(\sigmamu^2 + k\del{\svec_{i\ctrol},\svec_{j\ctrol}} + \delta_{ij} \sigman^2\) \\
        \hline
    \end{tabular}
    \egroup
    \caption{
        Shorthand notation for covariance matrices. The spatial coordinates of the \(i^\mathrm{th}\) treatment unit are denoted by \(\svec_{i\treat}\),
and those of the \(j^\mathrm{th}\) control unit by \(\svec_{j\ctrol}\), while \(\sentinel_i\) denotes the \(i^\mathrm{th}\) sentinel location along the border.
        \label{table:notation}
    }
\end{table}

\section{Estimating Linear Regression Coefficients}
\label{sec:betahat}

We present the posterior mean of the linear regression coefficients vector \(\betavec\) for the model specified in \autoref{eq:covariates_model}.
\begin{equation}
    \begin{split}
        &
        \cov\del{\Yvec \mid \betavec }
                = \begin{bmatrix}
                    \STT & 0 \\
                    0 & \SCC
                  \end{bmatrix}
            \,,\ 
            \cov\del{\betavec} = \sigmabeta^2 \eye_p 
            \,,\  
            \cov\del{\Yvec, \betavec} = \sigmabeta^2 \Dmat \\&
        \SigmaMat_Y = \cov\del{\Yvec} 
            = \cov\del{\Yvec \mid \betavec}
                + \sigmabeta^2 \Dmat \Dmat\trans \\&
        \hat\betavec 
            = \E\del{\beta \mid \Yvec} 
            = \cov\del{\betavec,\Yvec} \cov\del{\Yvec,\Yvec}^{-1} \Yvec
            = \sigmabeta^2 \Dmat\trans\SigmaMat_{Y}^{-1} \Yvec
    \end{split}
\end{equation}
The treatment and control residuals can then be obtained as \(\residvec = \Yvec - \Dmat \hat\betavec\).
Conditionally on \(\betavec=\hat\betavec\), the residuals of the treatment and control units then have independent multivariate normal distributions with covariances \(\STT\) and \(\SCC\) respectively.

\section{Calibration of Inverse-variance Test}
\label{sec:calibration}

We seek to obtain a valid hypothesis test against the null hypothesis of zero treatment effect everywhere along the border by using the inverse-variance weighted LATE estimate obtained in \autoref{sec:invvar} as a test statistic.

Under the parametric null hypothesis \(\modnull\), \(\yt\) and \(\yc\) are drawn from a single Gaussian process, with no discontinuity at the border.
Their joint covariance is
\begin{equation}
\cov \del{\begin{pmatrix}\yt \\ \yc\end{pmatrix} \mid \modnull } 
    = \begin{bmatrix}
        \STT & \KTC \\
        \KTC \trans & \SCC
    \end{bmatrix}\,,
\end{equation}
where \(\KTC\) is the \(n_\treat \times n_\ctrol\) matrix with \(ij^{\mathrm{th}}\) entry equal to \(k\del{\svec_{i\treat},\svec_{j\ctrol}}\).
The predicted mean outcomes \autoref{eq:postvar2gp_t_or_c} at the sentinels \(\muvec_{\sentinels \mid T}\) and \(\muvec_{\sentinels \mid T}\) are obtained by left-multiplying \(\yt\) and \(\yc\) by matrices \(\WT\) and \(\WC\) (respectively) that are deterministic functions of the unit locations and the hyperparameters:
\begin{equation}
    \WT = \KBT \STT^{-1} \quad\text{and}\quad
    \WC = \KBC \SCC^{-1}\,.
\end{equation}

Under \(\modnull\), the joint distribution of \(\muvec_{\sentinels \mid T}\) and \(\muvec_{\sentinels \mid T}\) is consequently also multivariate normal with mean zero and covariance given by:
\begin{equation}
\cov \del{\begin{pmatrix}\WT \yt \\ \WC \yc \end{pmatrix} \mid \modnull } = \begin{bmatrix}
    \WT \STT       \WT\trans & \WT \KTC \WC\trans \\
    \WC \KTC\trans \WT\trans & \WC \SCC \WC\trans
\end{bmatrix}\,.
\end{equation}
Continuing in this fashion, the cliff height \autoref{eq:postvar2gp} estimate 
\(\muvec_{\sentinels \mid Y} = \WT \yt - \WC \yc\)
is yet another zero-mean multivariate normal with covariance given by:
\begin{equation}
        \cov \del{\muvec_{\sentinels \mid Y} \mid \modnull} 
        = \WT \STT \WT\trans + \WC \SCC \WC\trans - \WT \KTC \WC\trans -  \WC\KTC\trans\WT\trans \,.
\end{equation}

Weighted LATE estimators of the form defined in \autoref{eq:weighted_estimator} are linear transformations of \(\muvec_{\sentinels \mid Y}\) and so under \(\modnull\), they are normally distributed with mean zero.
For a given weight function \(\weightb\), its variance is given by
\begin{equation}
        \var\del{\mu_{\tauw \mid Y} \mid \modnull} = \cov\del{ \frac{\weightb(\sentinels)\trans \muvec_{\sentinels \mid Y}}{ \weightb(\sentinels)\trans\ones_{\numsent}}}
        = \frac{\weightb(\sentinels) \trans \cov \del{\muvec_{\sentinels \mid Y}} \weightb(\sentinels)}{\del{ \weightb(\sentinels)\trans\ones_{\numsent}}^2}
        \,.
\end{equation}

The \(p\)-value follows from treating the LATE estimate as a test statistic.
Under the null hypothesis, the probability of \(\mu_{\tauw \mid Y}\) exceeding in magnitude its observed value \(\mu_{\tauw \mid Y^{obs}}\) is:
\begin{equation}
    \Pr\del{ \abs{\mu_{\tauw \mid Y}} \ge \abs{\mu_{\tauw \mid Y^{obs}}} \mid \modnull} = 2 \Phi\del{ -\abs{\mu_{\tauw \mid Y^{obs}}} \big/ {\sqrt{\var({\mu_{\tauw \mid Y} \mid \modnull})}} }\,.
\end{equation}
The calibrated inverse-variance test of \autoref{sec:hypothesis_testing} is the special case of this procedure with weights \(\weightb(\sentinels) = \Sigma^{-1}_{\sentinels \mid Y} \ones_{\numsent}\).

\end{appendices}

\pagebreak
\begin{center}
\phantomsection
	\addcontentsline{toc}{section}{Supplementary Materials}
	\bf 
	\LARGE
	Supplementary Materials
	\\*
	\large
	\georddtitle
\end{center}
\setcounter{equation}{0}
\setcounter{figure}{0}
\setcounter{table}{0}
\setcounter{section}{0}
\newcommand{\sprefix}{S-}
\renewcommand{\theequation}{\sprefix\arabic{equation}}
\renewcommand{\thesection}{\sprefix\arabic{section}}
\renewcommand{\thefigure}{\sprefix\arabic{figure}}
\renewcommand{\thetable}{\sprefix\arabic{table}}

\section{Spatial Confounding of Projected 1D RDD}
\label{sec:confounding}

\begin{figure}[tb]
    \centering
    \includegraphics[height=0.3\textheight]{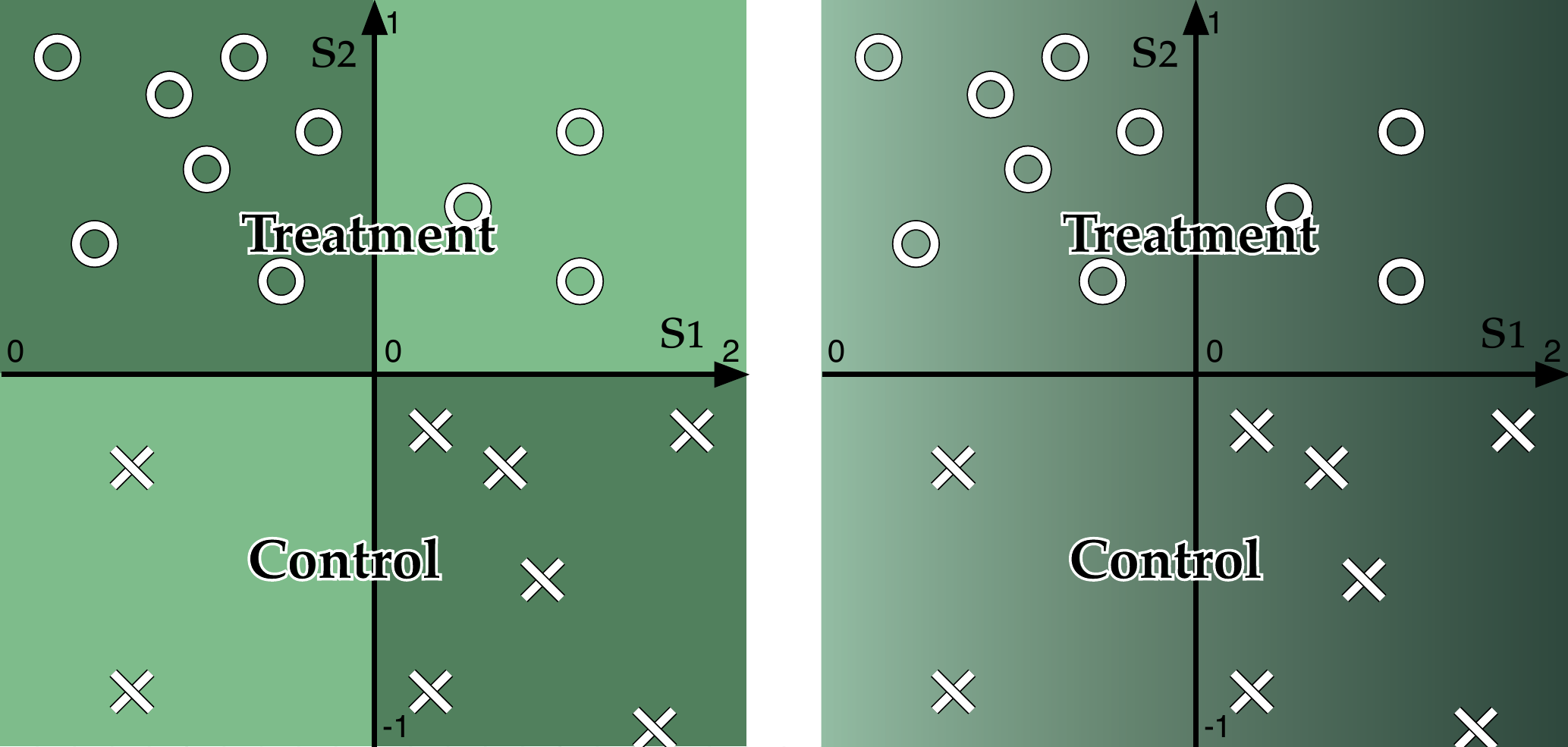}
    \caption{
        A theoretical example illustrating the susceptibility of the projected 1D~RDD method to spatial confounding. 
        On the left, the background shows the density of units, while on the right it shows the linear spatial trend in outcomes.
        The locations of treatment and control units are shown with circles and crosses respectively, separated by a border at \(\svec_2 = 0\). 
        Notice how treatment units are densest in an area with low outcomes, while control units units are densist in an area with high outcomes.
        \label{fig:confounding}}
\end{figure}

	Analysing GeoRDDs by using the signed distance from the border as a forcing variable in a 1D~RDD can lead to spatial confounding.
We demonstrate this with a simple artificial example, depicted in \autoref{fig:confounding}.
Suppose we have units in a 2D square, with spatial coordinates \(\svec_1 \in [0,2]\), and \(\svec_2 \in [-1,1]\), and with a straight border at \(\svec_2=0\) separating a treatment region from a control region.
Let us impose the null hypothesis, with outcomes driven only by a linear spatial trend running parallel to the border:
\begin{equation}
    Y_{i} = \alpha \svec_{1i} + \epsilon_i \,,\quad\epsilon_i \sim \normal\del{0, \sigman^2}
    \,.
\end{equation}
Let us consider the situation where the density \(\rho(\svec)\) of units is different in each quadrant of the square:
\begin{equation}
    \begin{aligned}
        \rho(\svec) = 2\rho_0 & \text{, where }\svec_1 < 1,~\svec_2 > 0 & \text{ (top left)} \\
        \rho(\svec) = \rho_0 & \text{, where }\svec_1 > 1,~\svec_2 > 0 & \text{ (top right)} \\
        \rho(\svec) = 2\rho_0 & \text{, where }\svec_1 > 1,~\svec_2 < 0 & \text{ (bottom right)}  \\
        \rho(\svec) = \rho_0 & \text{, where }\svec_1 < 1,~\svec_2 < 0 & \text{ (bottom left)}
    \end{aligned}
\end{equation}
The projection RDD then considers a 1D RDD along \(\svec_2\).
The usual RDD estimand \autorefexternal{eq:rdd_univ_estimand} can be obtained analytically:
\begin{equation}
    \tau = \frac{
            \int_{0}^{1}  2\rho \alpha \svec_{1} \dif \svec_{1} 
            + \int_{1}^{2}  \rho \alpha \svec_{1} \dif \svec_{1}
        }{
            \int_{0}^{1}  2\rho \dif \svec_{1} 
            + \int_{1}^{2}  \rho \dif \svec_{1}
        }
        -
        \frac{
            \int_{0}^{1} \rho \alpha \svec_{1} \dif \svec_{1} 
            + \int_{1}^{2} 2\rho \alpha \svec_{1} \dif \svec_{1}
        }{
            \int_{-1}^{0} \rho \dif \svec_{1} 
            + \int_{0}^{1} 2\rho \dif \svec_{1}
        }
    = \frac{-\alpha}{3}
    \,,
\end{equation}
and is non-zero even though the treatment effect is zero everywhere along the border.
This is because \(\svec_1\) acts as a hidden confounder whose distribution changes discontinuously at the border, which leads to bias and inconsistency in the projected 1D~RDD estimate.
In geographical settings, a discontinuous change in the density of units at the border is not unusual: for example a border could run alongside a park or a body of water, giving zero population density on one side of the border.
A visual inspection of \autorefexternal{fig:sales_map} showing the locations of units in a New York City property sales dataset reveals examples of this.

\section{Additional LATE Estimands and Simulations}
\label{sec:additional_late}

	In the paper, we presented and characterized four choices of local average treatment effect (LATE) estimands (and corresponding estimators): the uniformly-weighted \(\unifavg\), population-weighted \(\taurho\), inverse-weighted \(\invvar\), and finite population projected \(\tauproj\).
We here present two other estimands of interest not directly presented in the paper, which extend the projection population idea of \(\tauproj\) to superpopulations.
We then compare all the estimators in a simulated demonstration to illustrate their differences.

\subsection{Projected Land LATE}
\begin{figure}[tb]
    \centering
    \includegraphics[width=\textwidth]{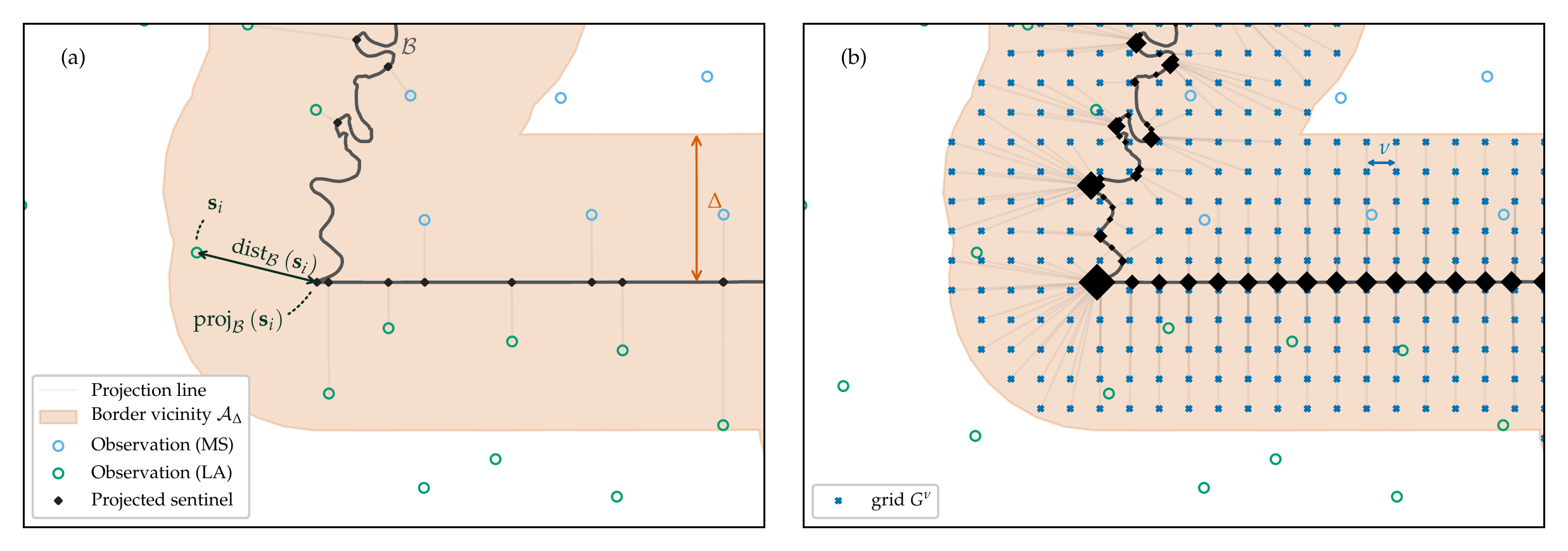}
    \caption{
		\label{fig:mississippi_projection_methods}
        Illustration of (a) projected finite-population LATE \(\tauproj\), and (b) projected land LATE \(\taugeo\), using the border separating Mississippi and Louisiana near Baton Rouge, with units at the centroid of each county.
        The border vicinity \(\area_\buffer\) is defined as all land within \(\buffer=50\,\mathrm{km}\) of the border.
    With both methods, every projected sentinel has equal weight in the LATE, but the tight grid in (b) causes sentinels to coincide or nearly coincide, which we depict by scaling up the size of the marker by the number of coinciding sentinels.}
\end{figure}

In certain applications, population-based estimands can be undesirable, especially if the locations at which measurements are made are not representative of the population of interest.
In such cases, geography-weighted estimands can be more natural.
See \cite{antonelli2016positive} for a discussion of this distinction in the context of preferential sampling.
Remember that the ``geometry-based'' estimand \(\unifavg\) places uniform weights along the border.
Instead, the ``geography-based'' projected land LATE estimand \(\taugeo\), illustrated in \autoref{fig:mississippi_projection_methods}(b), begins by placing uniform weights on the treatment and control areas \(\area_\treat\) and \(\area_\ctrol\) that are within distance \(\buffer\) of the border \(\border\), but then projects them onto the border to derive border weights.
In other words, the projection method from \(\tauproj\) is applied to an infinite population of uniform density on both sides of the border, instead of the finite population of observed units.

	We denote the border vicinity area by \(\area_\buffer\), defined as all points \(\svec\) such that \(\svec \in {\area_\treat \cup \area_\ctrol}\), and \(\dist_{\border}\del{\svec} < \buffer\).
To estimate \(\taugeo\), a tight grid \(\grid\) of evenly spaced points separated by \(\gridres\) is first generated covering \(\area_\buffer\).
Denote the number of grid points by \(L_\gridres\).
Each point \(\grid_l\), \(l=1,\dotsc,L_\gridres\) in \(\grid\) is then projected onto the border to become a sentinel.
The treatment effect at these positions is then estimated as before, yielding a mean vector and covariance matrix akin to \autorefexternal{eq:postvar2gp}.
The mean of the mean vector then gives an estimate of \(\taugeo\).
In other words, \(\taugeo\) is estimated by applying the \(\unifavg\) procedure with sentinels obtained by projecting the grid points, instead of equispaced sentinels.
\(\taugeo\) remains in the category of weighted-mean estimands, with the weight function \(\weightb(\sentinel)\) in \autorefexternal{eq:weighted_estimand} proportional to the area of \(\treatarea\) and \(\ctrolarea\) that \(\sentinel\) is nearest to, which can be written as the limit as the grid spacing goes to zero of point masses at the grid locations projected onto the border:
\begin{equation}
    \weightb(\sentinel) = \lim_{\gridres \rightarrow 0}\frac{1}{L_\gridres}\sum_{l=1}^{L_\gridres} \delta\del{\sentinel - \proj_{\border}\del{\grid_l} }\,.
\end{equation}

	For certain applications, it may be desirable to further restrict \(\area_\buffer\) to only certain types of land, for example residential areas in social studies, or farmland in agricultural studies.
It is important to note that \(\taugeo\) is never interpretable as the average treatment effect in the vicinity of the border, that is \(\taugeo \neq \int_{\area_\buffer} \tau(\svec) \dif \svec\).
Estimating the latter estimand would require predicting the conditional regression function at grid locations within the treatment or control region using only observations on the \emph{other} side of the border, which increases the extent of extrapolation required and thus makes the analysis more vulnerable to model misspecification.

\subsection{Projected Super-Population LATE}

	The purely geographical estimand \(\taugeo\) can be modified by weighting the grid points \(\grid_l\), \(l=1,\dotsc,L_\buffer\) by the population density \(\rho(\grid_l)\).
This gives the projected superpopulation LATE \(\taupop\).
Similarly to the density-weighted LATE \(\taurho\), estimating \(\taupop\) requires an estimate of the density \(\rho(\grid_l)\) at every grid point.
As before, the uncertainty in the estimate of \(\rho\) should in principle be propagated to the estimate of \(\taupop\), which generally will make the posterior distribution of \(\taupop\) neither normal nor analytically tractable.

	The estimand \(\taupop\) can be interpreted as giving equal weight to each unit in the superpopulation of units within the border vicinity \(\area_\buffer\), but then moving each unit from its original location to the nearest point on the border (where the GeoRDD is best able to estimate the treatment effect without undue extrapolation) and then averaging the treatment effect of each unit in this displaced superpopulation.
The resulting weight function is:
\begin{equation}
    \weightb(\sentinel) = 
		\lim_{\gridres \rightarrow 0}
		\frac{1}{L_\gridres}
		\sum_{l=1}^{L_\gridres}
			\delta\del{\sentinel - \proj_{\border}\del{\grid_l} } 
			\rho(\grid_l) 
		\,.
\end{equation}

\subsection{Wiggly Border Simulation}
\label{sec:wiggly_border}

\begin{figure}[tb]
\centering
\includegraphics[height=0.35\textheight]{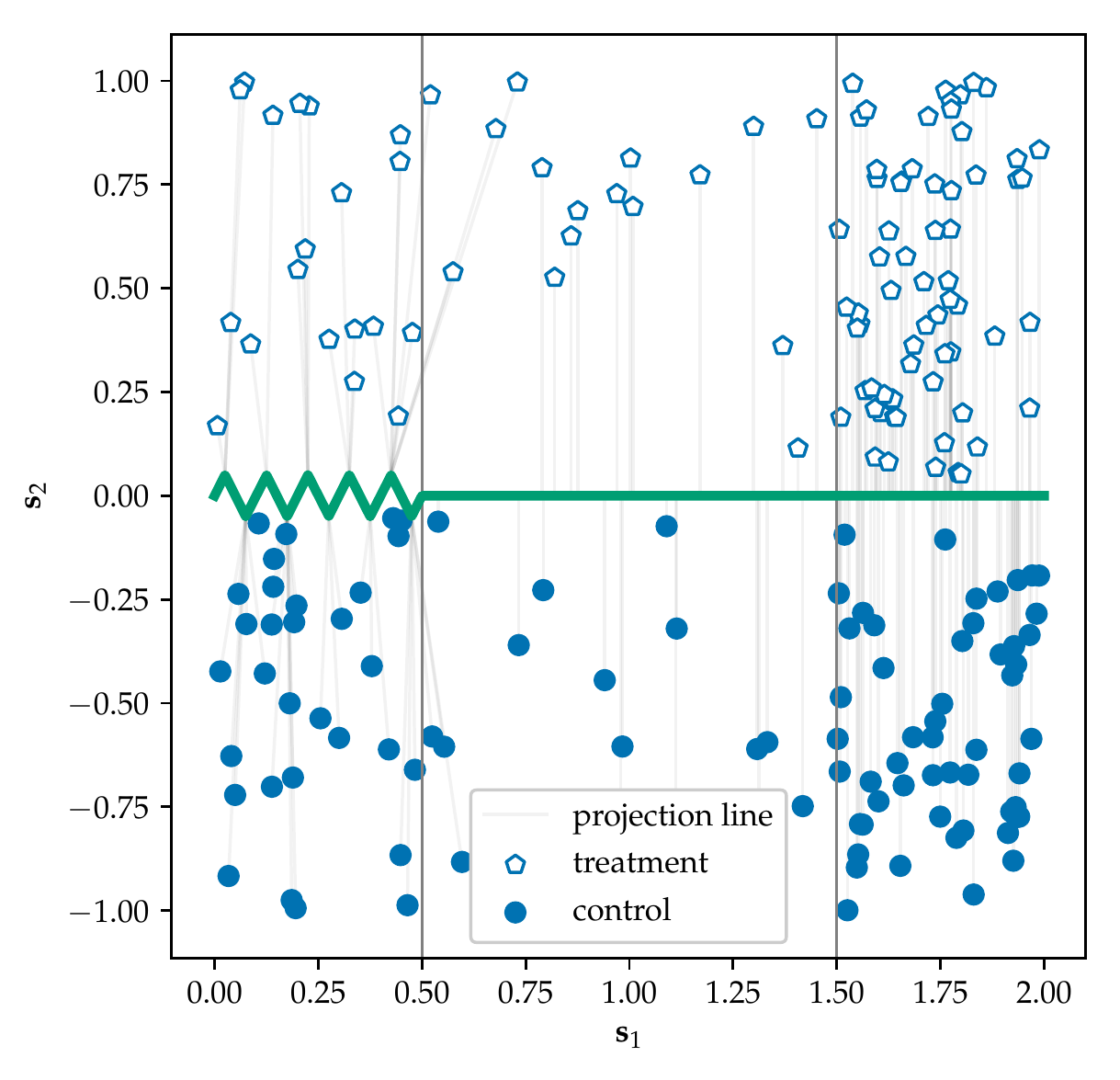}
\caption{
	\label{fig:wiggly_boundaries_setup}
	Spatial positions of units and border for the wiggly border simulation of \autoref{sec:wiggly_border}. Projection lines for the projected finite population LATE are shown in light gray.}
\end{figure}

	We illustrate all of our LATE estimators with a simulation.
200 units are placed in a square area delimited by spatial coordinates \(\svec_1 \in \cbr{0,2}\) and \(\svec_2 \in \cbr{-1, 1}\).
A border at \(\svec_2=0\) divides units vertically into a control and treatment region,
which are then further divided horizontally at \(\svec_1=0.5\) and \(\svec_1=1.5\) into three blocks:
\begin{itemize}
\item
  The leftmost block \(\svec_1 < 0.5\) has a weak treatment effect and density defined to be equal to \(\rho=1\).
\item
  The middle block \(0.5 \ge \svec_1 < 1.5\) has a much lower population density \(\rho=0.3\), and a stronger treatment effect.
\item
  The rightmost block \(\svec_1 \ge 1.5\), has a much higher population density \(\rho=2\), and a very strong treatment effect.
\end{itemize}
Furthermore, the border in the leftmost block is a triangular wave, to create ``wiggliness.''
We increase the number of wiggles from 0 to 1000 to observe the effect on the estimates.
The simulation setting is summarized in \autoref{table:wiggly_setup}.
We draw a single set of spatial coordinates, shown in \autoref{fig:wiggly_boundaries_setup}, then draw 10,000 simulations of the outcomes \(Y\) from a Gaussian process with squared exponential kernel (\(\ell=0.4\), \(\sigma=0.5\)).
To units above the border we add a treatment effect \(\tau(\svec_1, \svec_2) = \svec_1\).

\begin{table}[tbp]
\centering
\bgroup
\def\arraystretch{1.1}
\begin{tabular}{llll}
\hline
& Left \(\svec_1< 0.5\) & Middle \(0.5 \ge \svec_1 < 1.5\) & Right \(1.5 \ge \svec_1\)\tabularnewline
\hline
\textbf{Border} & wiggly & straight & straight\tabularnewline
\textbf{Density} & \(\rho=1\) & very low \(\rho=0.3\) & high \(\rho=2\)\tabularnewline
\(\taubold\) & weak & medium & strong\tabularnewline
\hline
\end{tabular}
\egroup
\caption{
	Summary of wiggly border simulation setup. 
	\label{table:wiggly_setup}}
\end{table}

\begin{figure}[!tb]
	\centering
	\includegraphics[height=0.6\textheight]{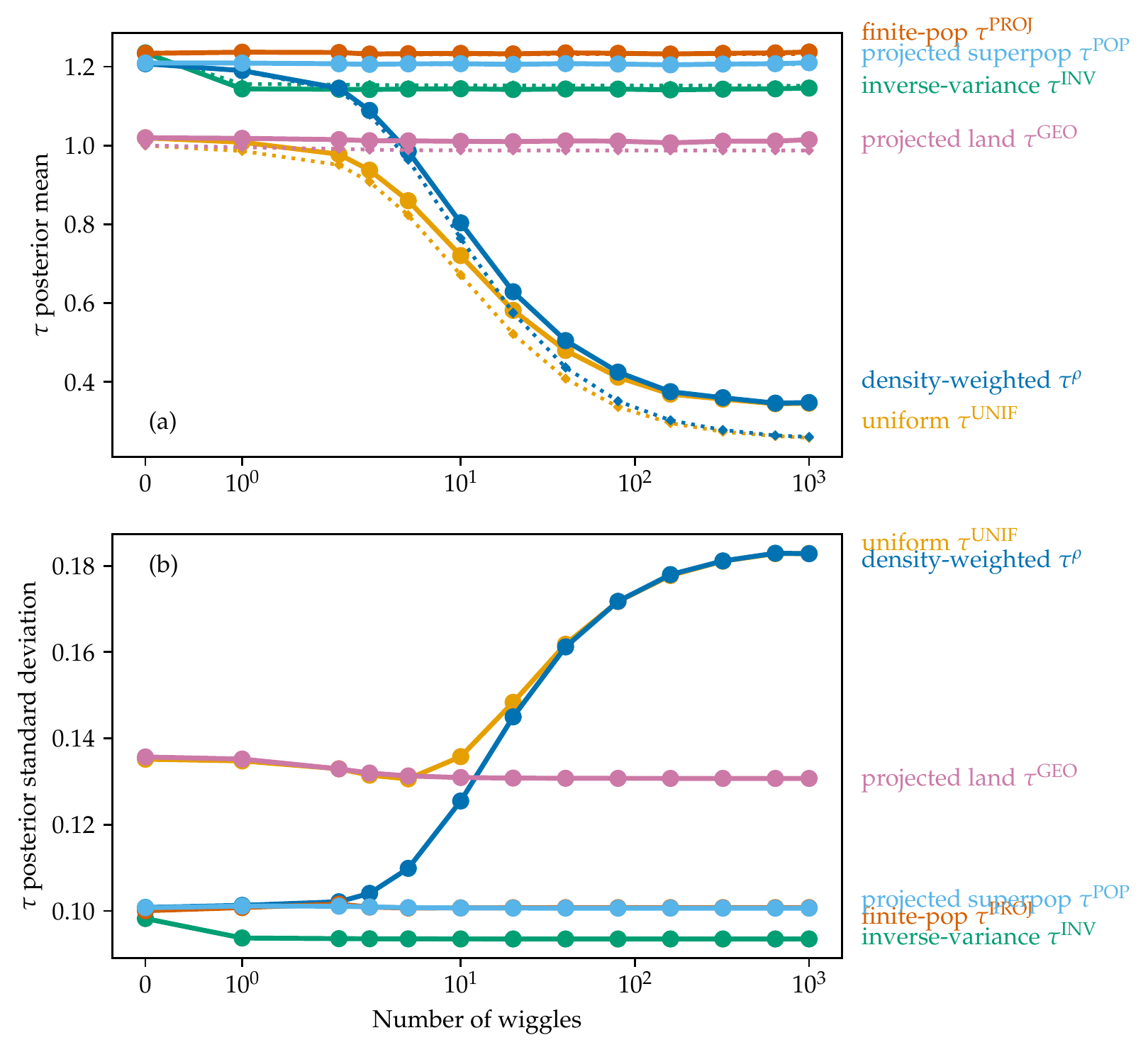}
	\caption{
		\label{fig:wiggly_boundaries_posteriors}
		Results of the simulations of \autoref{sec:wiggly_border}, showing for each LATE estimator as the leftmost section of the border gets wigglier (a) the estimate (posterior mean) averaged over 10,000 simulations with the corresponding estimand shown as a dotted line of the same color, and (b) the posterior standard deviation.
	}
\end{figure}

	We fit the Gaussian process model \autorefexternal{eq:spec2gp},
using the known hyperparameters of the covariance kernel and a weak prior on the mean parameter of each region,
and estimate the LATE using the six methods proposed above.
For projection-based methods, the buffer distance \(\buffer\) is infinite, so all of \(\area\) is included.
In \autoref{fig:wiggly_boundaries_posteriors}(a) we show, for each estimator, the corresponding estimand and average posterior mean estimate evolving as the number of border wiggles increases.
The behavior of the posterior standard deviation is shown in \autoref{fig:wiggly_boundaries_posteriors}(b).
The simulations results can also be found in \autoref{table:wiggly_results}.

	When the border is a straight line and because \(\treatarea\) and \(\ctrolarea\) are rectangles, the density-weighted estimand \(\taurho\) equals the projected superpopulation estimand \(\taupop\).
They are in fact both equal to the infinite-population average treatment effect since the treatment effect does not depend on \(\svec_2\).
Correspondingly, the posteriors of \(\taurho\) and \(\taupop\) are identical.
\(\taupop\) and the finite-population projected LATE \(\tauproj\) are also similar, but the latter has the advantage of not requiring local estimates of the population density.

	The geometry- and geography-based LATE \(\unifavg\) and \(\taugeo\) are also equivalent when the border is a straight line.
They give equal weight to the sparsely populated middle band, which produces a lower estimate with higher variance than the posteriors of \(\taurho\) and \(\taupop\).

	Lastly, the information-based inverse-variance estimand \(\invvar\) does not exactly coincide with any others.
The estimand and mean estimate change slightly from 0 to 1 wiggles, but remains stable thereafter, demonstrating the robustness of this estimator to border topology.
Weighting by the inverse variance gives the lowest posterior variance within the class of LATEs under consideration, which can indeed be seen in \autoref{fig:wiggly_boundaries_posteriors}(b).

	As we introduce wiggles into the leftmost band,
\(\taurho\) and \(\unifavg\) show their susceptibility to the border topology.
Proportionally more sentinels are packed into the leftmost section of the border,
upweighting the lower treatment effect of that band,
and resulting in a drop of the two estimates and estimands.
Meanwhile, \(\invvar\) remains stable despite the wiggles,
because the additional sentinels in the leftmost
band get automatically downweighted as their correlation rises.
The estimators that rely on projection
\(\tauproj\), \(\taugeo\), and \(\taupop\) also remain stable,
because the projected sentinels hardly move.
    
\begin{figure}[ptb]
\centering
\includegraphics[width=\textwidth]{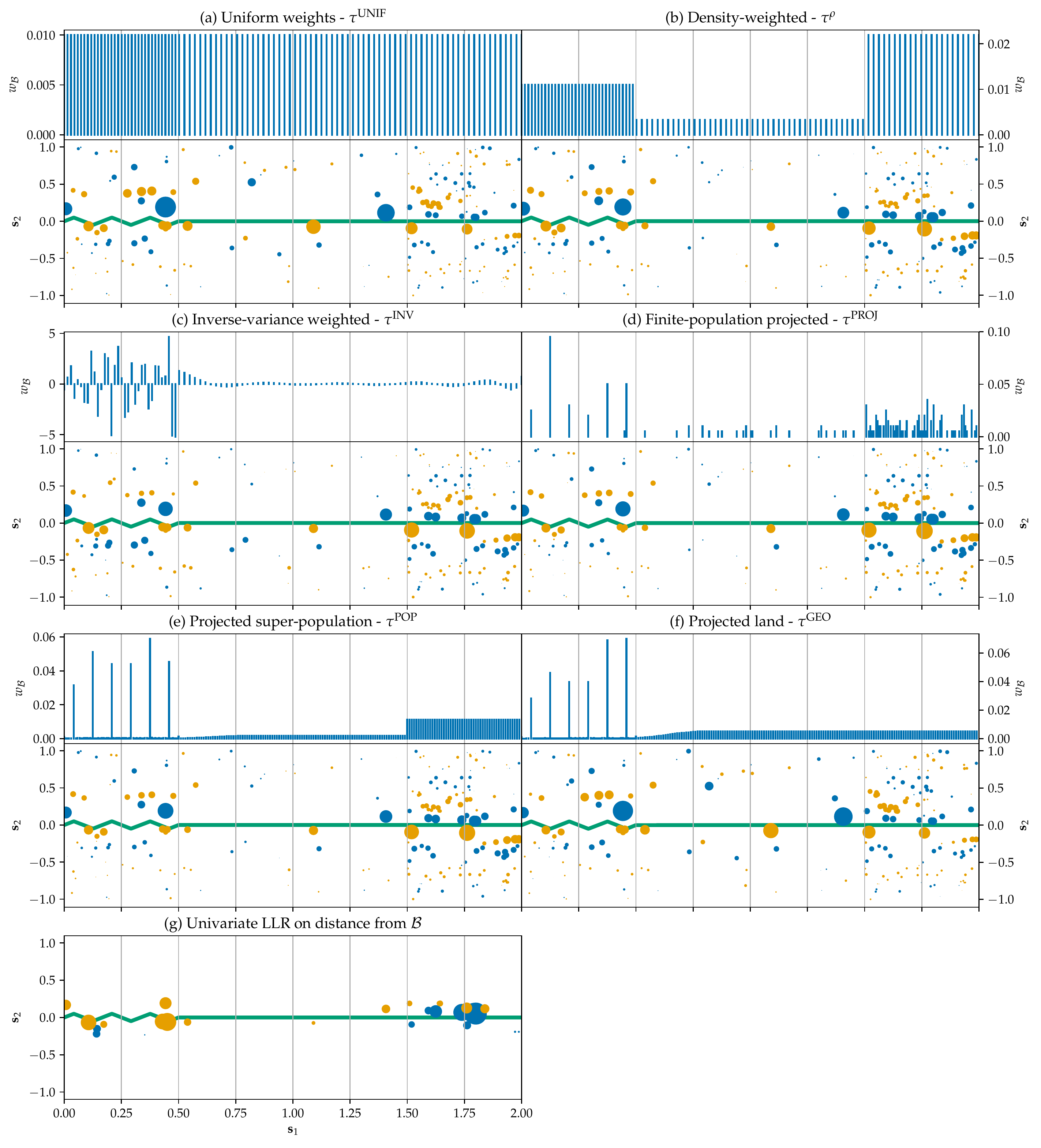}
\caption{
	\label{fig:weight_functions}
	Weight functions and induced weights on the observations for the six weight functions proposed in this paper. The weight function plots show the weight \(\weightb(\sentinel)\) against each sentinel's \(\svec_1\) coordinate. Sentinels with coinciding or nearly coinciding (within 0.005 of each other) coordinate \(\svec_1\) were merged and their weights summed. The induced weight plots show a circle for each unit, with the area of the circle proportional to its weight (\(\wt\) and \(\wc\) for treatment and control units respectively), and colored in blue for positive weights and orange for negative weights.}
\end{figure}

	In \autoref{fig:weight_functions}(a-f), we illustrate the behavior of border weights \(\weightb(\sentinel)\) and unit weights (\(\wt\) and \(\wc\)) in this simulation setting with 3 wiggles.
Note how evenly spaced sentinels (for \(\unifavg\), \(\taurho\), and \(\invvar\)) are more densely packed along \(\svec_1\) in the leftmost area because of the zig-zagging border.
The inverse-variance weighted estimator border weights can be seen to respond to this change in the border topology, though it is difficult to interpret their oscillating behavior.
While these border-weights look unreasonable and unstable, the induced unit weights for \(\invvar\) are well-behaved, and in fact quite similar to those of the projected finite- and infinite-population estimators.
Furthermore, note that all estimators can give some small negative weights \(\wt\) to treatment units, and small positive weights \(\wc\) to control units.
For Gaussian processes, this can be understood in terms of the negative side-lobes of the equivalent kernel (see \cite{rasmussen2006gaussian} Section 2.6).
The high variance of \(\unifavg\) and \(\taugeo\) manifests itself as large weights given to a small number of units.
All other estimators spread the weights more evenly amongst the units near the border, which reduces their variance.
    
	For comparison, the weights placed on units by the projected 1D RDD are shown in \autoref{fig:weight_functions}(g).
A triangular kernel in \(\svec_2\) was used with bandwidth selected using the MSE-minimizing method proposed by \cite{imbens2012optimal}.
The Projected 1D~RDD estimator can also be written as a linear combination of the observed outcomes \autorefexternal{eq:unit_weights}, and the unit weight vectors can be derived as:
\begin{equation}
\wt = \Xmat_b (\Xmat_\treat\trans \Wmat_\treat \Xmat_\treat)^{-1} \Xmat_\treat\trans \Wmat_\treat 
\quad\text{and}\quad
\wc = - \Xmat_b (\Xmat_\ctrol\trans \Wmat_\ctrol \Xmat_\ctrol)^{-1} \Xmat_\ctrol\trans \Wmat_\ctrol \,, 
\label{eq:unit_weights_llr}
\end{equation}
where \(\Xmat_b = \del{1~0}\), \(\Xmat_\treat\) is the \(n_\treat \times 2\) design matrix with the first column filled with ones and the second column containing the distance from the border of each treatment unit, and \(\Wmat_\treat\) is an \(n_\treat \times n_\treat\) diagonal matrix where the \(i^\mathrm{th}\) diagonal element is the triangular kernel evaluated on the \(i^\mathrm{th}\) unit's distance from the border.
The \(\Xmat_\ctrol\) and \(\Wmat_\ctrol\) matrices are analogously defined for control units.
By construction, the unit weights drop to zero outside of the support of the kernel.
Within the support, Projected 1D RDD can also give negative weights to treatment units, and positive weights to control units.
This results from the negative influence on the prediction \(\widehat{y^*}\) at \(x^*\) that univariate linear regression can give to an observation \(Y_i\) at \(X_i\) sufficiently far away on the opposite side of the mean \(\overline{X}\) of all observations.
Strikingly, almost all of the positive weights are given to units in the rightmost treatment area that are closest to the border, and almost all the negative weights are given to units in the leftmost control area.
Consequently, any trend in the outcomes across \(\svec_1\) would confound the estimated treatment effect.

\begin{landscape}
	\begin{table}[p]
    	\begin{tabular}{r|rrrrrrrrrrrr}
        	\hline
        	\(n_{\mathrm{wiggles}}\) & \(\widehat{\unifavg}\) & \(\unifavg\) & \(\widehat{\invvar}\) & \(\invvar\) & \(\widehat{\taurho}\) & \(\taurho\) & \(\widehat{\tauproj}\) & \(\tauproj\) & \(\widehat{\taugeo}\) & \(\taugeo\) & \(\widehat{\taupop}\) & \(\taupop\)\\
        	\hline
			0 & 1.02 (0.14) & 1.00 & 1.24 (0.10) & 1.23 & 1.21 (0.10) & 1.21 & 1.23 (0.10) & 1.24 & 1.02 (0.14) & 1.00 & 1.21 (0.10) & 1.21 \\
			1 & 1.01 (0.13) & 0.99 & 1.14 (0.09) & 1.16 & 1.19 (0.10) & 1.19 & 1.24 (0.10) & 1.24 & 1.02 (0.14) & 1.00 & 1.21 (0.10) & 1.21 \\
			2 & 0.98 (0.13) & 0.95 & 1.14 (0.09) & 1.15 & 1.15 (0.10) & 1.14 & 1.24 (0.10) & 1.24 & 1.01 (0.13) & 0.99 & 1.21 (0.10) & 1.21 \\
			3 & 0.94 (0.13) & 0.91 & 1.14 (0.09) & 1.15 & 1.09 (0.10) & 1.08 & 1.23 (0.10) & 1.23 & 1.01 (0.13) & 0.99 & 1.21 (0.10) & 1.21 \\
			5 & 0.86 (0.13) & 0.82 & 1.14 (0.09) & 1.15 & 0.98 (0.11) & 0.96 & 1.23 (0.10) & 1.23 & 1.01 (0.13) & 0.99 & 1.21 (0.10) & 1.21 \\
			10 & 0.72 (0.14) & 0.67 & 1.14 (0.09) & 1.15 & 0.80 (0.13) & 0.76 & 1.23 (0.10) & 1.23 & 1.01 (0.13) & 0.99 & 1.21 (0.10) & 1.21 \\
			20 & 0.58 (0.15) & 0.52 & 1.14 (0.09) & 1.15 & 0.63 (0.14) & 0.58 & 1.23 (0.10) & 1.23 & 1.01 (0.13) & 0.99 & 1.21 (0.10) & 1.21 \\
			40 & 0.48 (0.16) & 0.41 & 1.14 (0.09) & 1.15 & 0.50 (0.16) & 0.44 & 1.23 (0.10) & 1.23 & 1.01 (0.13) & 0.99 & 1.21 (0.10) & 1.21 \\
			80 & 0.41 (0.17) & 0.34 & 1.14 (0.09) & 1.15 & 0.42 (0.17) & 0.35 & 1.23 (0.10) & 1.23 & 1.01 (0.13) & 0.99 & 1.21 (0.10) & 1.21 \\
			160 & 0.37 (0.18) & 0.30 & 1.14 (0.09) & 1.15 & 0.38 (0.18) & 0.30 & 1.23 (0.10) & 1.23 & 1.01 (0.13) & 0.99 & 1.20 (0.10) & 1.21 \\
			320 & 0.36 (0.18) & 0.27 & 1.14 (0.09) & 1.15 & 0.36 (0.18) & 0.28 & 1.23 (0.10) & 1.23 & 1.01 (0.13) & 0.99 & 1.21 (0.10) & 1.21 \\
			640 & 0.34 (0.18) & 0.26 & 1.14 (0.09) & 1.15 & 0.35 (0.18) & 0.26 & 1.23 (0.10) & 1.23 & 1.01 (0.13) & 0.99 & 1.21 (0.10) & 1.21 \\
			1000 & 0.35 (0.18) & 0.26 & 1.15 (0.09) & 1.15 & 0.35 (0.18) & 0.26 & 1.24 (0.10) & 1.23 & 1.01 (0.13) & 0.99 & 1.21 (0.10) & 1.21 \\
			\hline
    	\end{tabular}
    	\caption{
        	\textbf{Wiggly Border Simulation Results.} 
        	Posterior mean averaged over 10,000 simulations, posterior standard deviation and true value for each LATE estimand as the wiggliness of the border is increased in the simulations of \autoref{sec:wiggly_border}.
        	\label{table:wiggly_results}
        	}
	\end{table}
\end{landscape}

\section{Alternate Tests for Non-Zero Treatment Effect}
\label{sec:alternate_tests}

	In our main paper, we present the calibrated inverse-variance test, which targets the weak null hypothesis of zero LATE.
It can be generalized to any other choice of LATE estimand defined as a weighted mean over the border, as in \autoref{eq:weighted_estimand}.
Tests of the sharp null hypothesis, that is \(\tau(\sentinel)=0\) for all \(\sentinel \in \border\), are also of interest, and we present two such tests in this section.
We provide a simulation comparing the power of the three tests when the treatment effect is constant, and apply each test to the NYC school district application of the main paper.
We advocate for the use of the calibrated inverse-variance test in most situations, as it has demonstrated higher power and robustness to model misspecification than the sharp null tests.

\subsection{Marginal Likelihood Test}

	Recall the null model \(\modnull\) defined in \autoref{sec:hypothesis_testing}; the unified Gaussian process is smooth and continuous at the border, and therefore accords with the sharp null hypothesis.
Intuitively, if there is a treatment effect, the likelihood of the observations should be lower under \(\modnull\) than under \(\modalt\),
the \(\GP\) model as specified in \autorefexternal{eq:spec2gp}.
We therefore choose the difference in log-likelihoods as our test statistic
\begin{equation}
    t = \log \Pr\del{\Yvec \mid \modalt} - \log \Pr\del{\Yvec \mid \modnull} \,,
\end{equation}
and wish to reject the sharp null hypothesis when its observed value \(t_{obs}\) is high.

	A parametric bootstrap approach is used to quantify what ``high'' means. We draw \(B\) bootstrap samples \(\Yvec^{(b)}\) from \(\modnull\),
using the same spatial locations as the original data,
and then fit the two competing models to the simulated data in order to obtain the bootstrapped test statistic
\begin{equation}
    t^{(b)} = \log \Pr\del{\Yvec^{(b)} \mid \modalt} - \log \Pr\del{\Yvec^{(b)} \mid \modnull}\,.
\end{equation}
Repeating this procedure, we obtain a distribution of \(t\) under \(\modnull\),
which we can then compare to the observed \(t\).
More precisely, the proportion of \(t^{(b)}\) drawn above \(t_{obs}\) estimates the \(p\)-value:
\begin{equation}
    p = \Pr\del{t > t_{obs} \mid \modnull}
                     \approx \frac{1}{B} \sum_{b=1}^B \Ind\cbr{t^{(b)} > t_{obs}}
                     \,.
\end{equation}
Computationally, because the hyperparameters and locations of the units are held constant during the bootstrap, we can reuse the Cholesky decomposition of the covariance matrix, allowing the test to be performed in seconds even with hundreds of units and thousands of bootstrap samples.

\subsection{``Chi-squared'' Test}
The likelihood-based sharp null above is valid and easy to understand.
But it may seem odd that the test aims to detect a non-zero treatment effect at the border, without any explicit reference to the border \(\border\).
The test statistic and \(p\)-values can be computed without access to the sentinel positions, using only the treatment and control indicators.
If the test is significant, there is no guarantee that this is due to a discontinuity at the border.

	To address this oddity, we can derive a test statistic directly from the cliff height estimator \autorefexternal{eq:postvar2gp}.
We use \(\muvec\) and \(\SigmaMat\) as shorthand for the posterior mean \(\muvec_{\sentinels \mid Y}\)
and covariance matrix \(\SigmaMat_{\sentinels \mid Y}\) throughout this section.
If a \(k\)-vector \(\yvec\) is distributed \(\normal\del{\muvec, \SigmaMat}\), with mean vector \(\muvec\) unknown and covariance \(\SigmaMat\) known, then under the null hypothesis that \(\mu=0\), the test statistic \(\yvec\trans \SigmaMat^{-1} \yvec\) has distribution \(\chi^2_k\).
See for example \cite{rencher2003methods} Section 5.2.2 for a classical derivation of this test.
This suggests that we could use \(S^2=\muvec\trans \SigmaMat^{-1} \muvec\) as a test statistic,
and obtain a \(p\)-value from a \(\chi^2_\numsent\) distribution function evaluated at \(S^2\), where \(\numsent\) is the number of sentinels.
However, we face two problems.
Firstly, this test, obtained heuristically from a Bayesian posterior by analogy with the classical multivariate normal result, is not a valid frequentist test.
Secondly, while \(\SigmaMat\) is mathematically full-rank, it is typically numerically rank-deficient.
Therefore, \(\numsent\) overestimates the true degrees of freedom of the null distribution.

	Benavoli and Mangili (2015), developing a test for function equality, address the second problem by trimming the \(\SigmaMat\) eigenvalues \(\lambda_i\) lower than \(\epsilon \sum_{j=1}^k \lambda_j\), with \(\epsilon\) a pre-specified small number (they use 0.01).
They address the first problem by showing that the resulting \(p\)-value is always conservative in their simulations.
However, in our work, we found the resulting \(p\)-value to be sensitive to the arbitrarily chosen \(\epsilon\) tolerance parameter, which makes it difficult to trust its validity.

	We therefore again take the parametric bootstrap approach, this time using \(S^2\) as the test statistic.
With B bootstrap samples, the \(p\)-value is estimated as
\begin{equation}
        p \approx \frac{1}{B} \sum_{b=1}^B \Ind\cbr{S_{(b)}^2 < S^2}
		\quad\text{with}\quad
        S_{(b)}^2 = (\muvec_{(b)} )\trans \SigmaMat^{-1} \muvec_{(b)}\,,
\end{equation}
where \(\muvec_{(b)}\) is the result of applying the cliff height estimator \autorefexternal{eq:postvar2gp} to the bootstrap sample \(\Yvec^{(b)}\).

	Because calculating \(S^2\) involves inverting a matrix \(\SigmaMat\) that is mathematically of full rank, but numerically of low rank, we may worry about the numerical stability of computing \(S\).
We verified in simulated examples that regularizing \(\SigmaMat\) by adding a small constant to its diagonal does not greatly affect the computed \(S^2\).
The parametric bootstrap ensures the frequentist validity of the test
regardless of the regularization.

\subsection{Comparing Power of Tests in Simulated Example}
\label{sec:powersim}

\begin{figure}[tb]
    \centering
    \includegraphics[height=0.4\textheight]{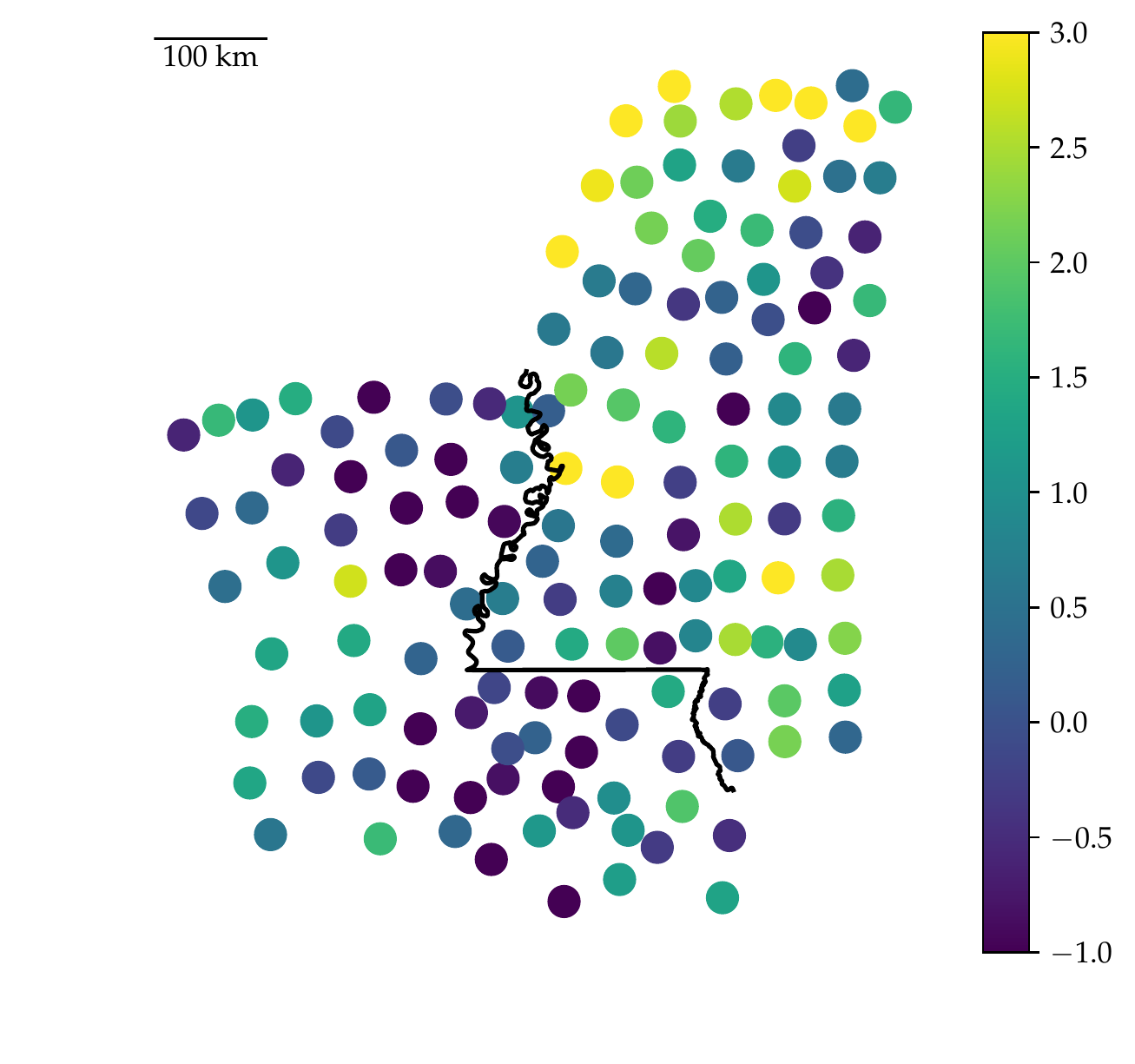}
    \caption{
		\label{fig:mississippi_counties}
		Set-up of the imaginary experiment in Louisiana and Mississippi. Each unit is at the centroid of a county. The colors indicated the observed outcomes in one draw of the simulation under \(\tau=1.5\). In this particular run, the \(p\)-values were 0.0016, 0.0018, and 0.0013 for the mLL, \(\chi^2\), and inverse-variance test respectively.}
\end{figure}

	The three tests we developed leverage different aspects of the design, and target two different null hypotheses. One may wonder how their power compares in the presence of a treatment effect. Considering once more the border between Louisiana and Mississippi, we imagine an experiment where the unit of analysis is the county, located at its centroid, as shown in \autoref{fig:mississippi_counties}.
We simulate outcomes from a single Gaussian process covering both states. For simplicity, we fix the hyperparameters to arbitrary values: \(\sigman=\sigmaf=1.0\) and \(\ell=100\,\mathrm{km}\).
We then add a constant treatment effect \(\tau\) to all the outcomes in Louisiana.
The results of the three tests proposed so far are shown in the first three rows of \autoref{table:power} for \(\tau=0\) (null hypothesis) and \(\tau=1.2\) and significance level \(\alpha=0.05\).

	We see that under the null, the \(\chi^2\) and likelihood ratio tests are valid (rejection of the null in 5\% of simulations up to simulation error).
This is enforced by the parametric bootstrap, which draws test statistics from the same null distribution to calibrate the tests.
However, the \(p\)-values for the inverse-variance test are biased down, so that we will falsely reject the null \(9\%\) instead of \(5\%\) of the time.
While unfortunate, this is unsurprising, since the inverse-variance test was derived heuristically rather than from a rigorous frequentist procedure.

\begin{table}
    \centering
    \bgroup
    \def\arraystretch{1.1}
    \centering
    \begin{tabular}{rrr}
        \hline
        & \multicolumn{2}{c}{Power under} \\
        Test & \(\tau=0\) & \(\tau=1.2\) \\
        \hline
	    Marginal log-likelihood bootstrap & 0.05 & 0.72 \\
	    \(\chi^2\) bootstrap & 0.05 & 0.63 \\
	    \(\invvar\) uncalibrated & 0.09 & 0.87 \\
	    \(\invvar\) calibrated & 0.05 & 0.80 \\
        \hline
    \end{tabular}
    \egroup
    \caption{
		Power of marginal likelihood, chi-squared, and inverse-variance tests, with nominal significance of \(\alpha=0.05\), under null and alternative hypothesis for simulated outcomes at the centroids of Louisiana and Mississippi counties.
    	\label{table:power}
	}
\end{table}

	After calibration, the hypothesis test based on the inverse-variance mean is valid, but retains higher power to detect the constant treatment effect than the mLL and \(\chi^2\) tests.
This can lead to a paradox: we may reject the weak null hypothesis, but fail to reject the sharp null hypothesis (using the \(\chi^2\) or likelihood test), even though rejection of the weak null logically implies rejection of the sharp null.
This paradox isn't specific to this setting, and is discussed in depth in the context of randomization-based inference by \cite{Ding:2014sf}.
To maximize power, we therefore recommend using the calibrated inverse-variance test in studies where the main interest is in the detection of an overall (average) increase or decrease in outcomes.

\subsection{Additional Tests for NYC School Districts Application}
\label{sec:nyc_hypothesis_tests}

\begin{table}[tbp]
    \centering
    \bgroup
    \def\arraystretch{1.1}
    \begin{tabular}{ll}
        \hline
        Test                   & \(p\)-value \\
        \hline
        Marginal log-likelihood bootstrap & 0.003     \\
        \(\chi^2\) bootstrap     & 0.022     \\
        \(\invvar\) uncalibrated & 0.0007    \\
        \(\invvar\) calibrated   & 0.002 \\
        \hline
    \end{tabular}
    \egroup
    \caption{
		Results of hypothesis tests for New York school district house prices.
		The marginal log-likelihood and \(\chi^2\) test were both performed with 10,000 bootstrap samples.
    	\label{table:NYC_tests}
	}
\end{table}

	We now compare the results of the three hypothesis tests applied to the NYC house prices application.
The three \(p\)-values are provided in \autoref{table:NYC_tests}, and show
agreement between the three tests, though the \(\chi^2\) test returns a considerably higher \(p\)-value, which is unsurprising considering the lower power of this test seen in simulations.

	To assess the validity of the three tests, we apply the placebo tests devised in \autorefexternal{sec:placebo}.
Within each district, we split the data in half by a line at angles \(1\degree\), \(3\degree\), \(5\degree\), \(6\degree\), \(\dotsc\), \(179\degree\).
Because these lines were drawn arbitrarily, we don't expect a discontinuous treatment effect between the two halves, and so we hope to see a uniform distribution of placebo \(p\)-values.
However, these tests will be highly correlated,
and so the low effective sample size could lead to some apparent departures from uniformity.
There is in fact visible autocorrelation in the graphs of placebo \(p\)-values as a function of angle.

\begin{figure}[tbp]
    \centering
    \includegraphics[width=\textwidth]{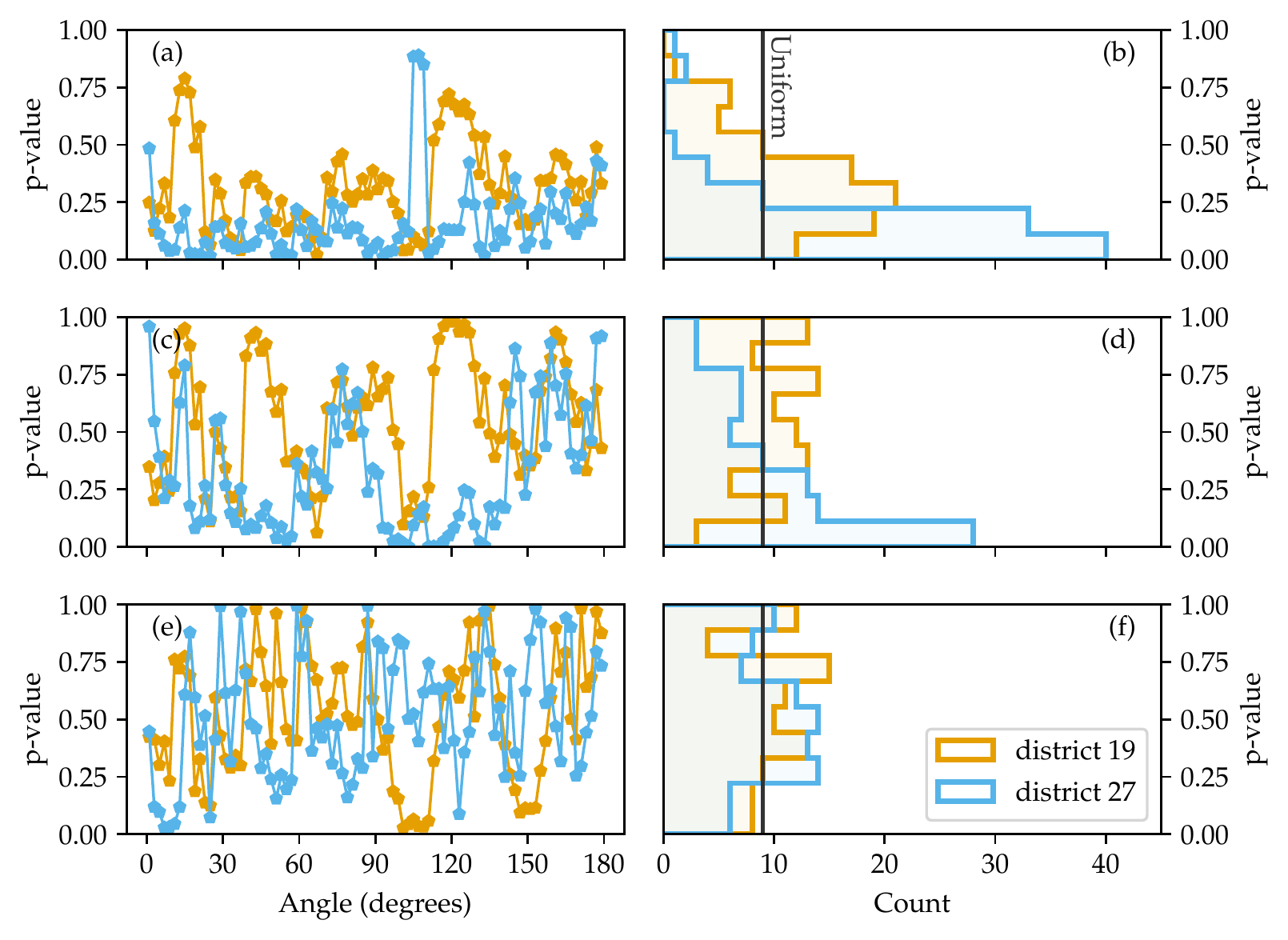}
    \caption{
		\label{fig:nyc_placebos}
		Placebo tests for significance tests applied to NYC school district house prices, applied within districts 19 and 27. From top to bottom, results are shown for the marginal log-likelihood bootstrap test, chi-squared bootrap test, and calibrated inverse-variance test. The first column shows the placebo \(p\)-value as a function of the border angle; the second column shows histograms of the placebo \(p\)-values, with the black vertical line indicating the uniform distribution.
	}
\end{figure}

	The mLL placebo \(p\)-values show a pronounced bias towards low values.
This seems to confirm our concern that the marginal log-likelihood may be sensitive to features of the data other than the discontinuity at the border.
In particular, model misspecification, which is a concern in spatial models, makes the interpretation of the mLL test unreliable.
Based on this vulnerability, and its manifestation in this application, we do not recommend relying on the likelihood-ratio test.

	The \(\chi^2\) test shows more robustness, with \autoref{fig:nyc_placebos}(d) showing some negative bias in district 27, and some positive bias in district 19, which could simply be due to the low effective sample size.
We therefore believe that the \(\chi^2\) test will continue to be reliable under misspecification.
It is only due to its low power that we hesitate to recommend its use in applications where the treatment effect is expected to be fairly homogenous.

	Lastly, the calibrated inverse-variance placebo \(p\)-values display no obvious bias, with \autoref{fig:nyc_placebos}(f) close to uniformly distributed, and \autoref{fig:nyc_placebos}(e) showing a lower auto-correlation than the mLL and \(\chi^2\) tests.
The high power and robustness of the inverse-variance test make a strong case for its use in most applications.

\section{Full Analysis: NYC School Districts}
\label{pairs-of-school-districts}

\begin{figure}[!tb]
    \centering
    \includegraphics[width=0.97\textwidth,height=0.8\textheight,keepaspectratio]{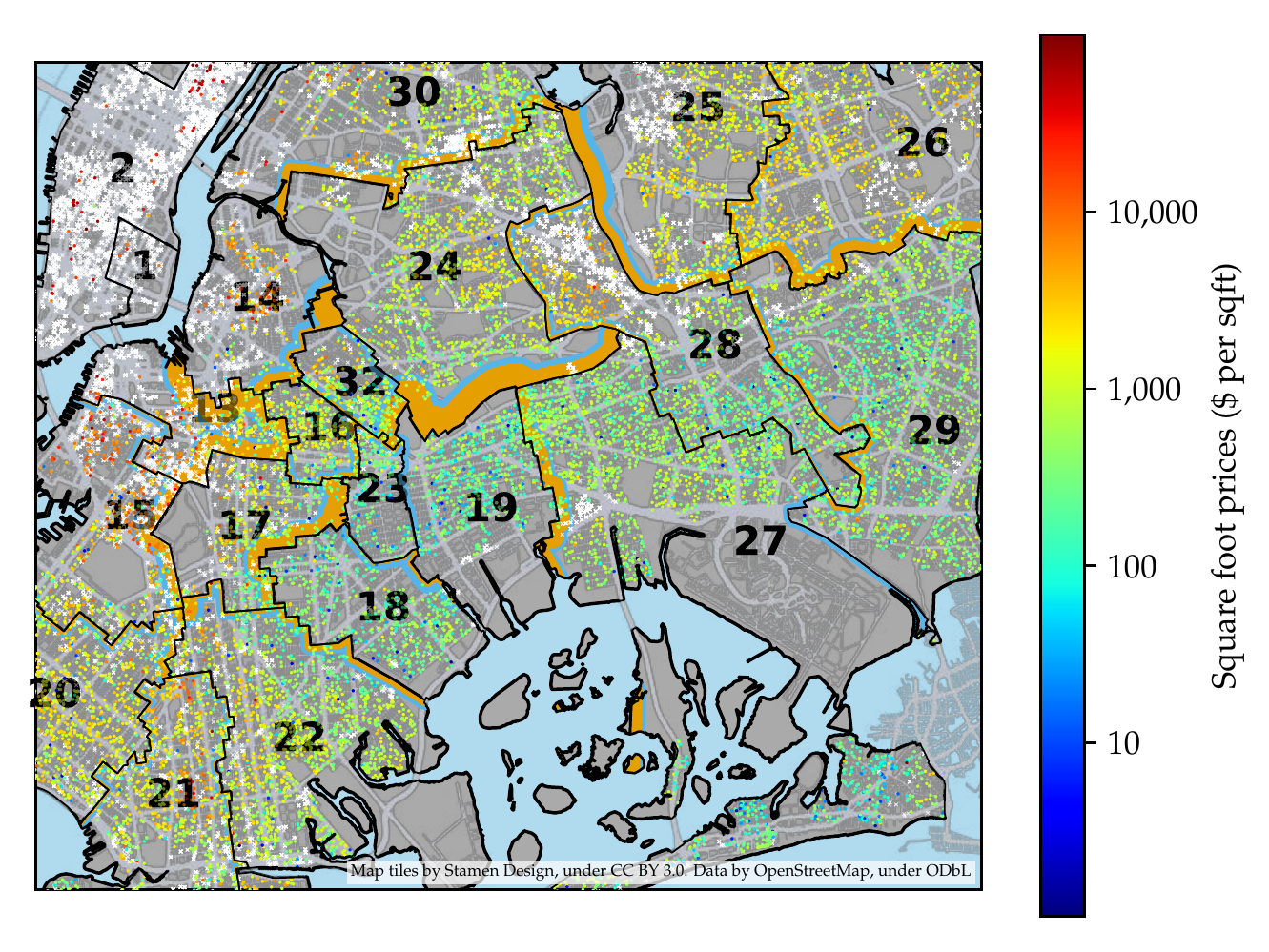}
    \caption{
		\label{fig:NYC_pairwise}
        Pairwise estimates of the inverse variance LATE between adjacent districts.
        The thickness of the orange buffer adjacent to borders is proportional to the posterior mean of the inverse variance LATE, and the blue buffer beyond it is proportional to the posterior standard deviation of the LATE.
    The buffers are drawn on the side of the border that is estimated to have higher house prices.}
\end{figure}

	The GeoRDD analysis can be repeated for each pair of adjacent districts.
\autoref{fig:NYC_pairwise} and \autoref{table:NYC_pairwise} give an overview of the results by showing the posterior mean and standard deviation of the inverse variance LATE estimated at each border.
Significant effects are found between many districts, but interpreting the results requires some caution.
We have already mentioned the issue of compound treatments for borders between school districts that overlap with the border between boroughs.
School districts 19, 32, and 14 are in Brooklyn, while districts 30, 24, and 27 are in Queens.

	Some school districts are separated by parks (or other non-residential zones), for example districts 15 \& 17 or 19 \& 24, so that house sales do not extend all the way to the border on one or both sides.
A significant treatment effect between these pairs cannot be interpreted as the detection of a discontinuity in prices at the border, let alone any kind of causal interpretation, but rather it means that the difference in prices between the two sides of the park exceeds the typical spatial variation of house prices expected over the same distance.
This is not unsurprising, and one may speculate that physical barriers like parks, rivers, railways and major roads can separate neighborhoods with distinct character, demographics and thus house prices.
This in turn challenges the stationarity assumption of the spatial model \autorefexternal{eq:spec2gp}.
The higher distance between data and the border also stretches the spatial model's ability to extrapolate, which makes it more vulnerable to model misspecification.

	Other pairs of district, like 13 \& 14, 13 \& 17, and 25 \& 28 have clusters of missing data (condo sales with unknown square footage) near the border that cast doubt on the interpretation of the estimated effect.
Nonetheless, significant effects are also found between pairs of school districts without issues due to compound treatments, physical barriers, or missing data.
House prices increase going across the border from districts 16 to 13, 18 to 17, 24 to 30, 23 to 17, 25 to 26, 28 to 29, and 29 to 26.
Overall, it seems that school district borders in Brooklyn and Queens can correspond to measurable jumps in house prices per square foot.
The estimated size of this effect varies: zero or negligible in some cases, such as between districts 15, 20, 21, and 22; and quite pronounced in others, such as a 20\% price increase from 29 to 26, or 22\% from 18 to 17.

\begin{landscape}
    \begin{table}[p]
        \footnotesize
        \begin{tabular}{r|lllllll}
            \hline
\( \mathbf{13} \)& \( \mathbf{14:}~-0.29 \pm 0.09 \)& \( \mathbf{15:}~+0.03 \pm 0.07 \)& \( \mathbf{16:}~+0.13 \pm 0.07 \)& \( \mathbf{17:}~+0.26 \pm 0.08 \)\\ 
\( \mathbf{14} \)& \( \mathbf{13:}~-0.29 \pm 0.09 \)& \( \mathbf{16:}~+0.16 \pm 0.10 \)& \( \mathbf{24:}~+0.38 \pm 0.15 \)& \( \mathbf{32:}~+0.07 \pm 0.12 \)\\ 
\( \mathbf{15} \)& \( \mathbf{13:}~+0.03 \pm 0.07 \)& \( \mathbf{17:}~+0.18 \pm 0.10 \)& \( \mathbf{20:}~-0.05 \pm 0.06 \)& \( \mathbf{22:}~-0.28 \pm 0.11 \)\\ 
\( \mathbf{16} \)& \( \mathbf{13:}~+0.13 \pm 0.07 \)& \( \mathbf{14:}~+0.16 \pm 0.10 \)& \( \mathbf{17:}~-0.04 \pm 0.07 \)& \( \mathbf{23:}~-0.10 \pm 0.07 \)& \( \mathbf{32:}~-0.05 \pm 0.06 \)\\ 
\( \mathbf{17} \)& \( \mathbf{13:}~+0.26 \pm 0.08 \)& \( \mathbf{15:}~+0.18 \pm 0.10 \)& \( \mathbf{16:}~-0.04 \pm 0.07 \)& \( \mathbf{18:}~+0.20 \pm 0.07 \)& \( \mathbf{22:}~+0.06 \pm 0.07 \)& \( \mathbf{23:}~-0.29 \pm 0.10 \)\\ 
\( \mathbf{18} \)& \( \mathbf{17:}~+0.20 \pm 0.07 \)& \( \mathbf{19:}~-0.06 \pm 0.12 \)& \( \mathbf{22:}~+0.10 \pm 0.07 \)& \( \mathbf{23:}~-0.03 \pm 0.09 \)\\ 
\( \mathbf{19} \)& \( \mathbf{18:}~-0.06 \pm 0.12 \)& \( \mathbf{23:}~-0.00 \pm 0.08 \)& \( \mathbf{24:}~+0.39 \pm 0.11 \)& \( \mathbf{27:}~+0.19 \pm 0.06 \)& \( \mathbf{32:}~-0.27 \pm 0.12 \)\\ 
\( \mathbf{20} \)& \( \mathbf{15:}~-0.05 \pm 0.06 \)& \( \mathbf{21:}~-0.04 \pm 0.05 \)& \( \mathbf{22:}~+0.11 \pm 0.08 \)\\ 
\( \mathbf{21} \)& \( \mathbf{20:}~-0.04 \pm 0.05 \)& \( \mathbf{22:}~-0.04 \pm 0.05 \)\\ 
\( \mathbf{22} \)& \( \mathbf{15:}~-0.28 \pm 0.11 \)& \( \mathbf{17:}~+0.06 \pm 0.07 \)& \( \mathbf{18:}~+0.10 \pm 0.07 \)& \( \mathbf{20:}~+0.11 \pm 0.08 \)& \( \mathbf{21:}~-0.04 \pm 0.05 \)\\ 
\( \mathbf{23} \)& \( \mathbf{16:}~-0.10 \pm 0.07 \)& \( \mathbf{17:}~-0.29 \pm 0.10 \)& \( \mathbf{18:}~-0.03 \pm 0.09 \)& \( \mathbf{19:}~-0.00 \pm 0.08 \)& \( \mathbf{32:}~+0.04 \pm 0.08 \)\\ 
\( \mathbf{24} \)& \( \mathbf{14:}~+0.38 \pm 0.15 \)& \( \mathbf{19:}~+0.39 \pm 0.11 \)& \( \mathbf{25:}~-0.26 \pm 0.13 \)& \( \mathbf{27:}~-0.22 \pm 0.10 \)& \( \mathbf{28:}~+0.06 \pm 0.06 \)& \( \mathbf{30:}~-0.14 \pm 0.05 \)& \( \mathbf{32:}~-0.02 \pm 0.08 \)\\ 
\( \mathbf{25} \)& \( \mathbf{24:}~-0.26 \pm 0.13 \)& \( \mathbf{26:}~+0.08 \pm 0.04 \)& \( \mathbf{28:}~-0.15 \pm 0.08 \)& \( \mathbf{29:}~-0.06 \pm 0.10 \)& \( \mathbf{30:}~+0.28 \pm 0.15 \)\\ 
\( \mathbf{26} \)& \( \mathbf{25:}~+0.08 \pm 0.04 \)& \( \mathbf{29:}~-0.18 \pm 0.05 \)\\ 
\( \mathbf{27} \)& \( \mathbf{19:}~+0.19 \pm 0.06 \)& \( \mathbf{24:}~-0.22 \pm 0.10 \)& \( \mathbf{28:}~-0.04 \pm 0.04 \)& \( \mathbf{29:}~+0.01 \pm 0.08 \)\\ 
\( \mathbf{28} \)& \( \mathbf{24:}~+0.06 \pm 0.06 \)& \( \mathbf{25:}~-0.15 \pm 0.08 \)& \( \mathbf{27:}~-0.04 \pm 0.04 \)& \( \mathbf{29:}~-0.09 \pm 0.04 \)\\ 
\( \mathbf{29} \)& \( \mathbf{25:}~-0.06 \pm 0.10 \)& \( \mathbf{26:}~-0.18 \pm 0.05 \)& \( \mathbf{27:}~+0.01 \pm 0.08 \)& \( \mathbf{28:}~-0.09 \pm 0.04 \)\\ 
\( \mathbf{30} \)& \( \mathbf{24:}~-0.14 \pm 0.05 \)& \( \mathbf{25:}~+0.28 \pm 0.15 \)\\ 
\( \mathbf{32} \)& \( \mathbf{14:}~+0.07 \pm 0.12 \)& \( \mathbf{16:}~-0.05 \pm 0.06 \)& \( \mathbf{19:}~-0.27 \pm 0.12 \)& \( \mathbf{23:}~+0.04 \pm 0.08 \)& \( \mathbf{24:}~-0.02 \pm 0.08 \)\\ 
            \hline
        \end{tabular}
        \caption{
            \label{table:NYC_pairwise}
            \textbf{Estimated Treatment Effects Between Adjacent NYC School Districts.}
            Each row gives the posterior (mean \(\pm\) standard deviation) of the inverse-variance LATEs for one district (row header) compared to its neighbors.
            For example the first cell indicates an estimated average change in log house prices per square foot of -0.29 when crossing the border from district 13 to 14.
        }
    \end{table}
\end{landscape}

\end{document}